\documentclass{ws-ijmpa}
\usepackage[super,compress]{cite}
\usepackage{graphicx}
\usepackage{url}

\usepackage{subfigure}
\usepackage{bbold}
\usepackage{dcolumn}
\usepackage{bm}
\usepackage[T1]{fontenc} 

\usepackage{braket}
\usepackage{multirow}
\usepackage{dsfont}
\usepackage{mathtools}
\usepackage{xcolor}

\newcommand{\I}{\mathrm{i}}
\newcommand{\Int}{\mathrm{int}}
\newcommand{\TO}{\rightarrow}

\newcommand{\diag}{\mathrm{diag}}
\newcommand{\INT}{\int_{\mathcal{R}^4} d^4 x\;}
\newcommand{\INTd}{\int_{\mathcal{R}^4} d^4 x^{\mathcal{D}}\;}
\newcommand{\gmu}{\gamma_{\mu}}
\newcommand{\g}[1]{\gamma_{#1}}
\newcommand{\Sl}{S(\Lambda)}
\newcommand{\St}{S(\Lambda)^{-1}}
\newcommand{\Par}{\mathcal{P}}
\newcommand{\Tem}{\mathcal{T}}
\newcommand{\Tr}{\mathrm{Tr}}
\newcommand{\PP}{\mathrm{P}}
\newcommand{\TT}{\mathrm{T}}
\newcommand{\PA}{\mathrm{PA}}
\newcommand{\TA}{\mathrm{TA}}
\newcommand{\RD}{R_{\DD}(\mathcal{D})}
\newcommand{\RDl}{R_{\DD}(\mathcal{D}^{\Lambda})}
\newcommand{\DD}{\mathrm{D}}
\newcommand{\UDalpha}{U_{\DD}^{\alpha}}
\newcommand{\Dlist}{\DD=\{\PP,\TT,\PA,\TA\}}
\newcommand{\ADe}{\mathcal{D}_{\mu\nu} A_{\nu}(\mathcal{D}x)}
\newcommand{\AD}{A_{\mu}(x)^{\mathcal{D}}}
\newcommand{\PSI}{\Psi^{\mathcal{D}}}
\newcommand{\PSIb}{\bar{\Psi}^{\mathcal{D}}}
\begin{document}
\markboth{Marco Catillo}{On $SU(2)_{CS}$-like groups and invariance of the fermionic action in QCD}

%
\catchline{}{}{}{}{}
%

\title{On $SU(2)_{CS}$-like groups and invariance of the fermionic action in QCD}

\author{Marco Catillo\footnote{
mcatillo@phys.ethz.ch}
}

\address{Institute for Theoretical Physics, ETH Z\"urich, 8093 Z\"urich, Switzerland}

\maketitle

\begin{history}
\received{Day Month Year}
\revised{Day Month Year}
\end{history}

\begin{abstract}
In this work, we introduce some new $U(1)$ symmetry groups of the free fermionic action in euclidean space-time, which are a consequence of parity and time-reversal symmetries. Afterwards, we discuss how the introduction of a gauge interaction affects the invariance of the action, with special reference to QCD. 
Moreover, inspired by recent QCD lattice results of \textit{Glozman et al.}, in which an interesting and unexpected symmetry group has been observed, namely $SU(2)_{CS}$ (that contains $U(1)_A$ as subgroup), we build other $SU(2)_{CS}$-like groups in euclidean space-time 
using the previous introduced $U(1)$ groups. 
Finally we argue about the possible invariance of the fermionic action with respect these new $SU(2)_{CS}$-like groups and its consequence on the hadron temporal correlators.

\keywords{Lattice QCD; Quantum Field Theory; High Temperature QCD}
\end{abstract}



\section{Introduction}\label{sec:Introduction}

As we mentioned in the abstract, the main motivation of this work comes from recent QCD lattice studies,\cite{Denissenya:2014poa,Denissenya:2014ywa,Denissenya:2015mqa,Denissenya:2015woa} in which unexpected symmetries emerge in the hadron spectrum. 
These new symmetries are manifested when the low-lying eigenmodes of the Dirac operator are subtracted, by hand, from the quark propagator in the calculation of hadron correlators.
The result of such studies is a degeneration of hadron masses, mesons and  baryons, connected via chiral and axial transformations. 
Moreover, it has been observed a larger and unexpected symmetry group among the hadron masses. 
This has been called $SU(2)_{CS}$, and contains the axial group as subgroup. 
The presence of $SU(2)_{CS}$ group symmetry, represents the extreme interesting novelty of these lattice calculations. 

Now, the emergence of $U(1)_A$, as well as chiral symmetry were quite expected in these studies.\cite{Denissenya:2014poa,Denissenya:2014ywa,Denissenya:2015mqa,Denissenya:2015woa} 
The reason is because the zero and the lowest eigenmodes of the Dirac operator are connected with the breaking of axial and chiral symmetry. 
In particular the zero modes break $U(1)_A$, this is the so called anomaly breaking.\cite{tHooft:1976snw} 
Instead, the lowest eigenmodes are connected with the breaking of chiral as well the axial symmetry. 
This is a consequence of the Banks-Casher relation,\cite{Banks:1979yr} 
which connects the chiral condensate $\langle\bar{\psi}\psi\rangle$ with the distribution $\rho(\lambda)$ of the lowest eigenvalues of the Dirac operator, i.e.  $\langle\bar{\psi}\psi\rangle = -\pi\rho(0)$. 
Therefore, it is understandable why the manual suppression of the lowest eigenmodes of the Dirac operator can lead to a restoration of chiral and $U(1)_A$ symmetry. 
However, the lowest eigenmodes seems to be also connected with the breaking of $SU(2)_{CS}$, which is a larger group than $U(1)_A$. 
This implies important consequences, which are given in Refs. \citen{Glozman:2017dfd,Glozman:2016ayd}. 
One of them is that at high temperature QCD, $T> T_c$ (where $T_c$ sets the restoration of chiral symmetry),  
where the lowest eigenmodes should be naturally suppressed (see the following lattice studies on this point,\cite{Bazavov:2012qja,Cossu:2013uua,Tomiya:2016jwr}) the $SU(2)_{CS}$ symmetry should emerge naturally. 
Studies on this direction has been made in Refs. \citen{Rohrhofer:2017grg,Rohrhofer:2019qwq,Rohrhofer:2019qal}. 
However, there are some recent results which are kind controversial, in the sense that $SU(2)_{CS}$ seems to disappear for $T\gg 3T_c$. 
Therefore only a particular range of temperatures, namely $T_c - 3T_c$ shows the $SU(2)_{CS}$ symmetry.

Another interesting feature of $SU(2)_{CS}$ is that it does not seem compatible with the deconfinement regime of QCD, because $SU(2)_{CS}$ as defined in Refs. \citen{Glozman:2017dfd,Glozman:2016ayd,Glozman:2015qva}, is not a symmetry of the action of free quarks (as we will see in section \ref{sec:old}), while it is the case for $U(1)_A$ and chiral group. 
That is perhaps the reason why at very high temperatures ($T\gg3T_c$) such symmetry disappears except for the subgroup $U(1)_A$, 
where still an effective restoration is observed.

Hence, the aim of this paper is to define $SU(2)_{CS}$-like groups containing the subgroup $U(1)_A$, and in such a way they are compatible with the presence of deconfinement, that can explain moreover the mass degeneracy found in Refs. \citen{Denissenya:2014poa,Denissenya:2014ywa,Denissenya:2015mqa,Denissenya:2015woa} and that can be in principle detected also for temperatures higher than $3T_c$. 
For doing so, we will proceed in our discussion using the euclidean formulation of QCD and we consider the following steps.  
At first, starting from the parity and time-reversal symmetries of QCD in euclidean space-time, we construct four $U(1)$ groups transformations of the spinor fields (section \ref{sec:u1}). 
Secondly, in section \ref{sec:fermionicaction}, we look at which conditions these $U(1)$ group transformations leave invariant the action of free fermions and we see what happens when we introduce a gauge field (in general non abelian) interaction. 
As third step, the introduction of such $U(1)$ groups will allow us to build two $SU(2)_{CS}$-like groups (which we name them $SU(2)_{CS}^{\mathcal{P}}$ and $SU(2)_{CS}^{\mathcal{T}}$), with a definition slight different from $SU(2)_{CS}$ introduced in Ref. \citen{Glozman:2015qva}. 
Moreover, we will show that they leave invariant the action of free massless quarks, as well we will discuss how the presence of a gauge field effects the breaking of these $SU(2)_{CS}$-like groups (section \ref{sec:su2cs}). In particular further properties of the $SU(2)_{CS}$-like groups are presented in section \ref{sec:inst}.
As last step, we will also see the consequences in hadron correlators of a possible $SU(2)_{CS}^{\mathcal{P}}$ symmetry in QCD. 
Finally we give our conclusions on where we summarize the main results of this paper.

However before looking the above points, we first remind in the next section, the original definition of $SU(2)_{CS}$ as given in Ref. \citen{Glozman:2015qva} in euclidean space-time, discussing why such definition is problematic if we want to make the action of free massless quarks invariant and consequently compatible with the phenomenon of deconfinement which we have at high temperatures in QCD.

\section{$SU(2)_{CS}$ group transformations}\label{sec:old}

In euclidean space-time the group $SU(2)_{CS}$ is defined by the following generators (see Ref. \citen{Glozman:2015qva}),

\begin{equation}
	\Sigma_i = \{\g{4},\I\g{5}\g{4},-\g{5}\},
	\label{eq:su2cs_generator}
\end{equation}

\noindent
where $\g{4}$ and $\g{5}$ are the usual gamma matrices (see the notation used in Eq. (\ref{eq:gamma})), and they satisfy the properties: $\{\g{4},\g{5}\}=0$ and $\g{4,5}^{\dagger}=\g{4,5}$, which implies that $\left[ \Sigma_i,\Sigma_j\right] = 2 \I\epsilon_{ijk}\Sigma_k$ and that $\Sigma_i^{\dagger} = \Sigma_i$, for all $i=1,2,3$. 
Moreover, from Eq. (\ref{eq:gamma}), we have $\Tr (\Sigma_i) = 0$ for all $i=1,2,3$, hence the set in Eq.  (\ref{eq:su2cs_generator}) forms a $su(2)$ Lie algebra.
Now a $SU(2)_{CS}$ transformation for a spinor field $\psi$ is given by

\begin{equation}
	SU(2)_{CS}:\,\psi(x) \TO\psi(x)^{(\Sigma)} = \exp(\I\alpha_n \Sigma_n)\psi(x),
	\label{eq:su2cs_psi}
\end{equation}

\noindent
where $(\alpha_1,\alpha_2,\alpha_3)$ is some given vector of global parameters. 
Regarding the $SU(2)_{CS}$ transformation of $\bar{\psi}$, this is not automatically fixed by Eq. (\ref{eq:su2cs_psi}). 
The reason is that in euclidean space-time, $\psi$ and $\bar{\psi}$ are independent. 
Therefore, taking different $SU(2)_{CS}$ transformations for $\psi$ and $\bar{\psi}$ is allowed, 
except that they are compatible with the hadron degeneracy found in Ref. \citen{Denissenya:2014poa,Denissenya:2014ywa,Denissenya:2015mqa,Denissenya:2015woa}. 
However we follow Ref. \citen{Glozman:2015qva} and we consider the  transformation for $\bar{\psi}$ as 

\begin{equation}
	SU(2)_{CS}:\,\bar{\psi}(x) \TO\bar{\psi}(x)^{(\Sigma)} = \bar{\psi}(x)\,\g{4}\, \exp(-\I\alpha_n \Sigma_n)\,\g{4},
	\label{eq:su2cs_barpsi}
\end{equation}

\noindent
which is how $\bar{\psi}$ would anyhow transform if we were in Minkowski space-time. 

In other lattice studies\cite{Rohrhofer:2017grg,Rohrhofer:2019qwq,Rohrhofer:2019qal}, other  ``versions" of $SU(2)_{CS}$ has been considered, just replacing the label $4\TO k$, with $k=1,2,3$ in Eqs. (\ref{eq:su2cs_generator}) to (\ref{eq:su2cs_barpsi}). 
However these changes bring to group transformations which are equivalent to our $SU(2)_{CS}$. 
The reason is due to the $\mathcal{O}(4)$ Lorentz symmetry of the QCD action in euclidean space-time. 
Indeed, as shown in Appendix \ref{app:B}, given a transformation $\Lambda\in \mathcal{O}(4)$ and its spinor representation $S(\Lambda)$, if we change the generators in Eq. (\ref{eq:su2cs_generator}) as $\Sigma_i \TO \Sigma_i^{\Lambda} = \Sl\Sigma_i \St$, 
then we obtain $SU(2)$ group transformations for $\psi$ and $\bar{\psi}$, 
let us say $SU(2)_{CS}^{\Lambda}$ ($\equiv SU(2)_{CS}\times \mathcal{O}(4)$), 
which are equivalent to $SU(2)_{CS}$ ones in Eqs. (\ref{eq:su2cs_psi}) and (\ref{eq:su2cs_barpsi}). 
In particular for $\Lambda =\bar{\Lambda}^k$, for $k=1,2,3$, with $\bar{\Lambda}$ given in Eq. (\ref{eq:lambar}), we get the generators of the new group, which are derived in Appendix \ref{app:B} (see Eq. (\ref{eq:su2cs_gen_k})), where we just need to substitute $\g{4}\TO\g{k}$ in Eqs. (\ref{eq:su2cs_generator}) to (\ref{eq:su2cs_barpsi}), up to some minus sign.

Unfortunately, although the transformations (\ref{eq:su2cs_psi})-(\ref{eq:su2cs_barpsi}) help to explain the degeneracy,\cite{Denissenya:2014poa,Denissenya:2014ywa,Denissenya:2015mqa,Denissenya:2015woa}
$SU(2)_{CS}$ is not a symmetry of the free quark action, as pointed out in Refs. \citen{Glozman:2015qva,Glozman:2017dfd,Glozman:2016ayd}. 
We can see this point, simply considering different direction of the vector $(\alpha_1,\alpha_2,\alpha_3)$ in (\ref{eq:su2cs_psi}) and (\ref{eq:su2cs_barpsi}), 
in which we can distinguish three important $U(1)$ subgroups of $SU(2)_{CS}$, 
that we will name $U(1)_{4}$, $U(1)_{4A}$ and the well-known axial group $U(1)_A$. 
We list their generator and group transformations for $\psi$ in Table \ref{tab:1}.

\begin{table}[tbh]
	\tbl{$U(1)$ subgroups of $SU(2)_{CS}$, obtained from Eqs. (\ref{eq:su2cs_psi}) and (\ref{eq:su2cs_barpsi}), taking different directions of the vector $(\alpha_1,\alpha_2,\alpha_3)$.}
		{\begin{tabular}{@{}cccc@{}}
				\toprule
			$U(1)\subset SU(2)_{CS}$ & Generator & Group trans. for $\psi$ & $\bm{\alpha}$\\
			\colrule
			$U(1)_{4}$ 	 & $\g{4}$			& $\exp(\I\alpha\g{4})$ & $(\alpha,0,0)$\\
			$U(1)_{4A}$ & $\I\g{5}\g{4}$	&  $\exp(\I\alpha(\I\g{5}\g{4}))$ & $(0,\alpha,0)$\\
			$U(1)_A$ 	 & $-\g{5} $		&  $\exp(-\I\alpha\g{5})$ & $(0,0,\alpha)$\\
			\botrule
	\end{tabular}\label{tab:1}}
\end{table}

Now, the groups $U(1)_{4}$ and $U(1)_{4A}$ are not symmetry groups of the massless quark action. 
This fact can be simply shown plugging the transformations in Eqs. (\ref{eq:su2cs_psi}) and (\ref{eq:su2cs_barpsi}) with the proper choice of $(\alpha_1,\alpha_2,\alpha_3)$ according to the Table \ref{tab:1}, in the action 

\begin{equation}
	S_F^0 (\psi,\bar{\psi}) = \INT \bar{\psi}(x) \g{\mu}\partial_{\mu}^x \psi(x), 
	\label{eq:sf0}
\end{equation}

\noindent
taking $\psi'(x) = U \psi(x)$ and $\bar{\psi}'(x) = \bar{\psi}(x)\,\g{4}U^{\dagger}\g{4}$, with $U\in U(1)_{4}$ or $U\in U(1)_{4A}$ and checking whether $S_F^0(\psi',\bar{\psi}') = S_F^0 (\psi,\bar{\psi})$. 
However this is not true. 
The reason is that writing $S_F^0 (\psi,\bar{\psi}) = \sum_{\mu}S_F^0 (\psi,\bar{\psi})_{\mu}$, with $S_F^0 (\psi,\bar{\psi})_{\mu} = \INT \bar{\psi}(x) \g{\mu}\partial_{\mu}^x \psi(x)$, with no sum over the repeated indices, 
then the terms $S_F^0 (\psi,\bar{\psi})_{\mu\neq4}$ are not invariant under $U(1)_{4}$ and $U(1)_{4A}$, because of the anticommutation properties of the gamma matrices (see them in Appendix \ref{app:A}). 

Therefore $S_F^0$ is not invariant under $SU(2)_{CS}$ because at least two of their subgroups (namely $U(1)_{4}$ and $U(1)_{4A}$) are broken explicitly. 
Moreover, the introduction of a mass term $S_m(\psi,\bar{\psi}) = m\INT \bar{\psi}(x)\psi(x)$ breaks explicitly $U(1)_A$ and $U(1)_{4A}$ as well. 

Hence there is no reason on observing such symmetry in the deconfinement regime of QCD, 
where quarks approach to almost free particles due to the weak behavior of the coupling constant. 

Therefore $SU(2)_{CS}$ has to be a symmetry on the region where confinement is still held, but as shown in numerous lattice calculations,\cite{Denissenya:2014poa,Denissenya:2014ywa,Denissenya:2015mqa,Denissenya:2015woa} 
it is a symmetry appearing in the regime when chiral symmetry is restored. 
The result is, therefore, having such symmetry in QCD in a small region of the phase diagram above chiral phase transition but not at too high temperature when deconfinement should occur. Look indeed the lattice studies in Refs. \citen{Rohrhofer:2017grg,Rohrhofer:2019qwq,Rohrhofer:2019qal}. 

Now, before trying to find possible solutions of all these issues of $SU(2)_{CS}$, 
we are going to discuss some other group transformations which will turn out to be useful for our purposes mentioned in the introduction.

\section{$U(1)$ groups from discrete symmetries}\label{sec:u1}

In this section, starting from parity and time-reversal symmetries of QCD in euclidean space-time, we build four interesting $U(1)$ group transformations of fermion fields. 
As we know, parity and time-reversal transformations of the space-time, are implemented as $x_{\mu}\TO x_{\mu}^{\Par} = \Par_{\mu\nu}\,x_{\nu}$ and $x_{\mu}\TO x_{\mu}^{\Tem} = \Tem_{\mu\nu}\,x_{\nu}$, where $\Par_{\mu\nu} = 2\delta_{\mu\nu}\delta_{\nu 4} - \delta_{\mu\nu}$ and $\Tem_{\mu\nu} = -\Par_{\mu\nu}$. 
Such transformations leave invariant the action of free fermions $S_F^0$, which implies that the fermion fields has to transform according to the proper representations of $\Par$ and $\Tem$. 
In euclidean space-time, such spinor transformations are given by\cite{Hernandez:2009zz}

\begin{equation}
	\begin{array}{ll}
		&\psi(x) \xrightarrow{\Par} \psi(x)^{\Par} = \g{4}\, \psi(\Par x),\\
		&\bar{\psi}(x) \xrightarrow{\Par} \bar{\psi}(x)^{\Par} =  \bar{\psi}(\Par x)\, \g{4},\\
		&\psi(x) \xrightarrow{\Tem} \psi(x)^{\Tem} = \I\g{4}\g{5}\, \psi(\Tem x),\\
		&\bar{\psi}(x) \xrightarrow{\Tem} \bar{\psi}(x)^{\Tem}= \bar{\psi}(\Tem x)\, \I\g{4}\g{5},\\
	\end{array}
	\label{eq:par_temp}
\end{equation}

\vspace{0.02cm}

\noindent
from which is straightforward to show that $S_F^0 (\psi,\bar{\psi}) = S_F^0 (\psi^{\Par},\bar{\psi}^{\Par}) = S_F^0 (\psi^{\Tem},\bar{\psi}^{\Tem})$. 
From Eq. (\ref{eq:par_temp}), it is worth to notice that applying two consecutive transformations of the spinor fields, we re-obtain the original spinors, i.e. 
$\psi(x)^{\Par^2} \equiv (\psi(x)^{\Par})^{\Par} = \psi(x)$ and 
$\psi(x)^{\Tem^2} \equiv (\psi(x)^{\Tem})^{\Tem} = \psi(x)$, 
(the same can be applied substituting $\psi\TO\bar{\psi}$ in the previous relations). 
This is because $\{\g{5},\g{4}\}=0$, $\g{4,5}^2 = \mathds{1}$ and $\mathcal{P}^2 = \mathcal{T}^2 = \mathcal{I}$. 
Therefore, if we apply $n$ times the discrete transformations in Eq. (\ref{eq:par_temp}), we get 

\begin{equation}
	\begin{split}
		&\psi(x)^{\Par^n} = \left\lbrace 
		\begin{array}{ll}
			\psi(x)&\mbox{for}\, n\,\mbox{even}\\
			\g{4}\,\psi(\Par x)&\mbox{for}\, n\,\mbox{odd}\\
		\end{array}
		\right.
		\\
		&\bar{\psi}(x)^{\Par^n} = \left\lbrace 
		\begin{array}{ll}
			\bar{\psi}(x)&\mbox{for}\, n\,\mbox{even}\\
			\bar{\psi}(\Par x)\,\g{4}&\mbox{for}\, n\,\mbox{odd}\\
		\end{array}
		\right.\\
		&\psi(x)^{\Tem^n} = \left\lbrace 
		\begin{array}{ll}
			\psi(x)&\mbox{for}\,n\,\mbox{even}\\
			\I\g{4}\g{5}\,\psi(\Tem x)&\mbox{for}\, n\,\mbox{odd}\\
		\end{array}
		\right.\\
		&\bar{\psi}(x)^{\Tem^n} = \left\lbrace 
		\begin{array}{ll}
			\bar{\psi}(x)&\mbox{for}\,n\,\mbox{even}\\
			\bar{\psi}(\Tem x)\,\I\g{4}\g{5}&\mbox{for}\, n\,\mbox{odd}\\
		\end{array}
		\right.
	\end{split}
	\label{eq:par_temp2}
\end{equation}

At this point we define the following transformations (the names on the left sides will be justified in a while)

	\begin{equation}
		\begin{split}
			U(1)_{\PP}:\,& \psi(x)\rightarrow\psi(x)^{U_{\PP}^{\alpha}}=
			\sum_{n=0}^{\infty}\frac{(\I \alpha)^n}{n!}(\psi(x))^{\Par^n},\\
			&\bar{\psi}(x)\rightarrow\bar{\psi}(x)^{U_{\PP}^{\alpha}}=
			\sum_{n=0}^{\infty}\frac{(-\I \alpha)^n}{n!}(\bar{\psi}(x))^{\Par^n},\\
			U(1)_{\TT}:\,& \psi(x)\rightarrow\psi(x)^{U_{\TT}^{\alpha}}=
			\sum_{n=0}^{\infty}\frac{(\I \alpha)^n}{n!}(\psi(x))^{\Tem^n},\\
			&\bar{\psi}(x)\rightarrow\bar{\psi}(x)^{U_{\TT}^{\alpha}}=
			\sum_{n=0}^{\infty}\frac{(-\I \alpha)^n}{n!}(\bar{\psi}(x))^{\Tem^n},\\
			U(1)_{\PA}:\,& \psi(x)\rightarrow\psi(x)^{U_{\PA}^{\alpha}}=
			\sum_{n=0}^{\infty}\frac{(\I \alpha)^n}{n!}(\I \g{5})^{k_n}(\psi(x))^{\Par^n},\\ &\bar{\psi}(x)\rightarrow\bar{\psi}(x)^{U_{\PA}^{\alpha}}=
			\sum_{n=0}^{\infty}\frac{(-\I \alpha)^n}{n!}(\I \g{5})^{k_n}(\bar{\psi}(x))^{\Par^n},\\
			U(1)_{\TA}:\, &\psi(x)\rightarrow\psi(x)^{U_{\TA}^{\alpha}}=
			\sum_{n=0}^{\infty}\frac{(\I \alpha)^n}{n!}(\I \g{5})^{k_n}(\psi(x))^{\Tem^n},\\ &\bar{\psi}(x)\rightarrow\bar{\psi}(x)^{U_{\TA}^{\alpha}}=
			\sum_{n=0}^{\infty}\frac{(-\I \alpha)^n}{n!}(\I \g{5})^{k_n}(\bar{\psi}(x))^{\Tem^n},\\
		\end{split}
		\label{eq:u1d_trans}
	\end{equation}

\noindent
where $k_n = 4 + (n \mod 2)$, $\alpha$ is some global real parameter and we have introduced the labels P, T, PA and TA only to distinguish such transformations. 
From Eq. (\ref{eq:u1d_trans}), we can split the sums for $n$ even and $n$ odd, and, exploiting Eq. (\ref{eq:par_temp2}), we can write them in a more compact way as 

\begin{equation}
	\begin{split}
		U(1)_{\DD}:\, &\psi(x)^{U^{\alpha}_{\DD}}= \cos (\alpha)\,\psi(x) +  \I\sin (\alpha)\, R_D (\mathcal{D}) \psi(\mathcal{D}x),\,\\
		&\bar{\psi}(x)^{U^{\alpha}_{\DD}}=\cos(\alpha)\,\bar{\psi}(x) -\I\eta\: \sin (\alpha)\, \bar{\psi}(\mathcal{D}x)R_D (\mathcal{D}),
	\end{split}
	\label{eq:u1d_trans2}
\end{equation}

\noindent
in which the label D can be one of the possible transformations, i.e. $\Dlist$. 
Instead, $\mathcal{D}$, $R_D (\mathcal{D})$ and the values of $\eta$ are reported in Table \ref{tab:2}, 
and they satisfy the properties: 

\begin{equation}
	\begin{split}
		&\RD = \RD^{\dagger},\quad\RD^2 = \mathds{1},\\
		&\mathcal{D}_{\mu\alpha}\mathcal{D}_{\alpha\nu} = \delta_{\mu\nu},\quad \det(\mathcal{D})=-1,
	\end{split}
	\label{eq:rdprop}
\end{equation}

\noindent
which are easy to check looking the definitions of the gamma matrices in Eq. (\ref{eq:gamma}) and the definition of parity $\mathcal{P}$ and time-reversal $\mathcal{T}$ matrices.

\begin{table}[htb]
		\tbl{Different values of D transformations and their representations. }
		{\begin{tabular}{@{}cccc@{}}\toprule
			D & $\mathcal{D}$ & $\RD$ & $\eta$ \\
			\colrule
			P & $\mathcal{P}$ & $\g{4}$ & $1$\\
			T & $\mathcal{T}$ & $\I\g{4}\g{5}$ & $1$ \\
			PA & $\mathcal{P}$ & $\I\g{5}\g{4}$ & $-1$ \\
			TA & $\mathcal{T}$ & $\g{4}$ & $-1$\\
			\botrule
		\end{tabular}
		\label{tab:2}}
\end{table}

Now, we show that the transformations given in Eq. (\ref{eq:u1d_trans2}) form $U(1)$ groups for all $\Dlist$. 
In order to do this we use the result of Appendix \ref{app:C}, namely

\begin{equation}
	\begin{split}
		&(\psi(x)^{\UDalpha})^{U_{\DD}^{\beta}} = \psi(x)^{U_{\DD}^{\alpha+\beta}},\\
		&(\bar{\psi}(x)^{\UDalpha})^{U_{\DD}^{\beta}} = \bar{\psi}(x)^{U_{\DD}^{\alpha+\beta}},
	\end{split}
	\label{eq:group_prop}
\end{equation}

\noindent
for $\Dlist$, 
where the prove is easy and it just makes use of Eqs. (\ref{eq:u1d_trans}) and (\ref{eq:u1d_trans2}) and the properties in (\ref{eq:rdprop}).
The Eq. (\ref{eq:group_prop}) shows the closure property of $U(1)_{\DD}$ transformations, i.e. that two consecutive $U(1)_{\DD}$ transformations give again a $U(1)_{\DD}$ transformation. 
Moreover, from Eq. (\ref{eq:group_prop}), we can see that such groups are also abelian, in fact $(\psi(x)^{U_{\DD}^{\alpha}})^{U_{\DD}^{\beta}} = 
\psi(x)^{U_{\DD}^{\alpha+\beta}} = (\psi(x)^{U_{\DD}^{\beta}})^{U_{\DD}^{\alpha}}$ 
and 
$(\bar{\psi}(x)^{U_{\DD}^{\alpha}})^{U_{\DD}^{\beta}} = 
\bar{\psi}(x)^{U_{\DD}^{\alpha+\beta}} = (\bar{\psi}(x)^{U_{\DD}^{\beta}})^{U_{\DD}^{\alpha}}$. 
Regarding the inverse of a generic element of $U(1)_{\DD}$, it can be obtained taking in Eq. (\ref{eq:u1d_trans2}) $(U_{\DD}^{\alpha})^{-1} = U_{\DD}^{-\alpha}$. 
Such group transformations contain obviously also the identity element which is given setting $\alpha = 0$ in Eq. (\ref{eq:u1d_trans2}). 
Therefore Eq. (\ref{eq:group_prop}) is enough to prove that $U(1)_{\DD}$ are abelian groups. 
In order to prove the unitarity of $U(1)_{\DD}$ transformations, we need to find a scalar product involving $\psi$ and $\bar{\psi}$ which is left invariant by the transformations (\ref{eq:u1d_trans2}). 
A simple scalar product can be $
(\psi_1,\psi_2) = \INT \psi_1(x)^{\dagger}\psi_2(x),
$ 
where $\psi_1$ and $\psi_2$ are two generic spinor fields. 
In Appendix \ref{app:D}, we prove that $(\psi_1^{U_{\DD}^{\alpha}},\psi_2^{U_{\DD}^{\alpha}}) = (\psi_1,\psi_2)$, 
and the same can be obtained with other pair of independent fields $ \bar{\psi}_1$ and $\bar{\psi}_2$, namely 
$(\bar{\psi}_1^{\dagger},\bar{\psi}_2^{\dagger}) = ((\bar{\psi}_1^{U_{\DD}^{\alpha}})^{\dagger},(\bar{\psi}_2^{U_{\DD}^{\alpha}})^{\dagger})$. 
It is important to notice that the scalar product $(\psi_1,\psi_2)$ is not a local function, since we are integrating over the all space-time. 
This is crucial in order to prove the unitarity of the $U(1)_{\DD}$ group transformations.
We end saying that since the abelian groups $U(1)_{\DD}$ are also unitary,  
this justifies the name that we have chosen so far for them.

Moreover, as we have done for $SU(2)_{CS}$, we can get other groups equivalent to $U(1)_{\DD}$ exploiting the $\mathcal{O}(4)$ Lorentz invariance of the free massless fermionic action in (\ref{eq:sf0}). 
This can be obtained replacing in Eq. (\ref{eq:u1d_trans2}), $\mathcal{D}\TO\mathcal{D}^{\Lambda} = \Lambda^{-1} \mathcal{D} \Lambda$, 
$R_{\DD}(\mathcal{D})\TO R_{\DD}(\mathcal{D}^{\Lambda})=S(\Lambda)R_{\DD}(\mathcal{D})S(\Lambda)^{-1}$, with $\Lambda\in\mathcal{O}(4)$, as we describe in Appendix \ref{app:B2}, look Eqs. (\ref{eq:u1d_c1}) and (\ref{eq:u1d_c2}).

\section{Fermionic action and $U(1)_{\DD}$ transformations}\label{sec:fermionicaction}

The fermionic action in QCD, that we give in Eq. (\ref{eq:sf}), can be split in the sum of different terms, 

\begin{equation}
	S_F(\psi,\bar{\psi},A) = S_F(\psi,\bar{\psi},A=0) + S_{\Int}(\psi,\bar{\psi},A)
	\label{eq:sf2}
\end{equation}

\noindent
with $S_F(\psi,\bar{\psi},A=0) = S_F^0 (\psi,\bar{\psi}) + S_m(\psi,\bar{\psi})$ and 
where $S_F^0$ is written in Eq. (\ref{eq:sf0}) and $S_m$ is the mass term, i.e. $S_m(\psi,\bar{\psi}) = m\INT\bar{\psi}(x)\psi(x)$, 
while 

\begin{equation}
	S_{\Int}(\psi,\bar{\psi},A) = \I\INT \bar{\psi}(x)\g{\mu}A_{\mu}(x) \psi(x),
	\label{eq:int}
\end{equation}

\noindent
with $A_{\mu}(x)$ the gauge field, which for our purposes we can consider it non-abelian. 

In section \ref{subsec:freeinv}, we want to show that $U(1)_{\PP}$ and $U(1)_{\TT}$ transformations leave invariant $S_F^0 + S_m$; 
$U(1)_{\PA}$ and $U(1)_{\TA}$ transformations leave invariant $S_F^0$, but not $S_m$. 
Finally we see (in section \ref{subsec:int}) how in general $S_{\Int}$ is not invariant under all $U(1)_{\DD}$ transformations. 
However we can find particular arrangements of the gauge field $A_{\mu}(x)$ with zero topological charge, which leaves $S_{\Int}$ invariant under $U(1)_{\DD}$ transformations, where $\Dlist$. 

It is important to say that in euclidean space-time the fermionic action in Eq. (\ref{eq:sf2}) is not the only part of the action involving the fermion fields and therefore subjected to the $U(1)_{\DD}$ transformations. 
Other terms in the action, in fact, arise from gauge fields in a non-zero topological sector (such as instantons). 
We are talking about the \textit{'t Hooft term} of the action that describes the interaction of fermion fields with the zero modes of the Dirac operator. 
However, since as it will be clear in section \ref{subsec:int}, the gauge fields with non-zero topological charge already can in principle break $U(1)_{\DD}$ invariance, therefore it is not worth to consider such \textit{'t Hooft term}, which comes from instanton gauge configurations, 
and that therefore breaks $U(1)_{\DD}$ by default.

\subsection{Invariance of the free fermion action}\label{subsec:freeinv}

Before starting our prove, we give the following useful relations involving $\RD$, $\mathcal{D}$ and $\eta$ of Table \ref{tab:2}, 

\begin{equation}
	\RD \g{\mu}\RD = \eta\mathcal{D}_{\mu\nu}\g{\nu},\qquad\mbox{and}\qquad
	\partial_{\mu}^x = \mathcal{D}_{\mu\nu}\partial_{\nu}^{\mathcal{D}x},
	\label{eq:rel1}
\end{equation}

\noindent
which are valid for all $\Dlist$ 
and they are pretty trivial to prove for each case listed in Table \ref{tab:2}.
Then, we see that under $U(1)_{\DD}$ transformations of the spinor fields, we get that $S_F(\psi,\bar{\psi},A=0)$ transforms as

\begin{equation}
	\begin{split}
		S_{F} (\psi^{U_{\DD}^{\alpha}},\bar{\psi}^{U_{\DD}^{\alpha}}, A = 0)&= 
		\INT\bar{\psi}(x)^{U_{\DD}^{\alpha}}(\g{\mu}\partial_{\mu}^x + m)\psi(x)^{U_{\DD}^{\alpha}}\\
		&=\cos(\alpha)^2\INT\bar{\psi}(x)(\g{\mu}\partial_{\mu}^x + m)\psi(x)\\ 
		&+ \I\cos(\alpha)\sin(\alpha)\INT\bar{\psi}(x)(\g{\mu}\partial_{\mu}^x+ m)\RD\psi(\mathcal{D}x)\\
		&-\I\eta\cos(\alpha)\sin(\alpha)\INT\bar{\psi} (\mathcal{D}x) \RD (\g{\mu}\partial_{\mu}^x + m)\psi(x)\\
		&+\eta\sin(\alpha)^2\INT\bar{\psi}(\mathcal{D}x) \RD(\g{\mu}\partial_{\mu}^x + m)\RD\psi(\mathcal{D}x) \\
		&=\cos (\alpha)^2\INT\bar{\psi}(x)(\g{\mu}\partial_{\mu}^x + m)\psi(x)\\
		&+ \I\cos(\alpha)\sin(\alpha)\INT\bar{\psi}(x)(\g{\mu}\partial_{\mu}^x + m)\RD\psi(\mathcal{D}x)\\
		&-\I  \eta  \cos(\alpha)\sin(\alpha)\INT\bar{\psi} (x)(\eta\:\g{\mu}\partial_{\mu}^x + m) \RD\psi(\mathcal{D}x)\\ &+\eta\sin(\alpha)^2\INT\bar{\psi}(x) (\eta\:\g{\mu}\partial_{\mu}^x + m)\psi(x),
	\end{split}
	\label{eq:sfree}
\end{equation}

\noindent
where in the 2nd equality we used Eq. (\ref{eq:u1d_trans2}) expanding all terms, 
in the 2nd and 3rd term of the last equality, we have changed the variable $x_{\mu}\TO x^{\mathcal{D}}_{\mu}=\mathcal{D}_{\mu\nu}x_{\nu}$, 
used Eq. (\ref{eq:rel1}) and the properties in Eq. (\ref{eq:rdprop}), 
so $d^4 x = \vert\det(\mathcal{D})\vert d^4 x^{\mathcal{D}} = d^4 x^{\mathcal{D}}$. 
Therefore for different values of $\eta$ (listed in Table \ref{tab:2}), we obtain

\begin{equation}
	\begin{array}{lll}
		\eta=1:\quad &S_{F} (\psi^{U_{\DD}^{\alpha}},\bar{\psi}^{U_{\DD}^{\alpha}}, A = 0) &= S_F(\psi,\bar{\psi},A=0)
		\\
		\eta=-1:\quad &S_{F} (\psi^{U_{\DD}^{\alpha}},\bar{\psi}^{U_{\DD}^{\alpha}}, A = 0) &= 
		\INT \bar{\psi}(x)\g{\mu}\partial_{\mu}^x \psi(x)\\
		& &+m \cos(2\alpha)\INT \bar{\psi}(x)\psi(x) \\
		& &+m\I\sin(2\alpha)\INT \bar{\psi}(x)\RD\psi(\mathcal{D}x). 
	\end{array}
	\label{eq:act_invd}
\end{equation}

\noindent
From Eq. (\ref{eq:act_invd}) is evident that the action of free massless ($m = 0$) fermions $S_F^0$ is $U(1)_{\DD}$-invariant for all $\Dlist$. 
However the mass term breaks $U(1)_{\PA}$ and $U(1)_{\TA}$, since in that case $\eta = -1$ from Table \ref{tab:2}. 
In conclusion, $S_F^0(\psi,\bar{\psi})+S_m(\psi,\bar{\psi})=S_F(\psi,\bar{\psi},A=0)$ is invariant under $U(1)_{\PP}$ and $U(1)_{\TT}$, and  $S_F^0(\psi,\bar{\psi})$ is invariant under $U(1)_{\PA}$ and $U(1)_{\TA}$ transformations.

\subsection{Gauge fields and $U(1)_{\DD}$ invariance}\label{subsec:int}

If now $A\neq0$, we need to consider the term $S_{\Int}$ in Eq. (\ref{eq:int}), in order to check if the full fermion action $S_F$ at $m=0$ is $U(1)_{\DD}$-invariant.
We define for convenience $\AD = \ADe$, with $\mathcal{D} = \mathcal{P}$ or $\mathcal{T}$, and we can see how $S_{\Int}$ transforms, i.e. 

\begin{equation}
	\begin{split}
		S_{\Int}(\psi^{U_{\DD}^{\alpha}},\bar{\psi}^{U_{\DD}^{\alpha}},A)&=
		\I\INT \bar{\psi}(x)^{U_{\DD}^{\alpha}} \g{\mu}A_{\mu}(x)\psi(x)^{U_{\DD}^{\alpha}}\\
		&=\I \cos(\alpha)^2\INT\bar{\psi}(x)\g{\mu}A_{\mu}(x)\psi(x)\\
		&-\sin(\alpha)\cos(\alpha)\INT \bar{\psi}(x)\g{\mu}A_{\mu}(x)\RD\psi(\mathcal{D}x)\\
		&+\eta\sin(\alpha)\cos(\alpha)\INT \bar{\psi}(x)(\mathcal{D}x)\RD\g{\mu}A_{\mu}(x)\psi(x)\\
		&+\I\eta\sin(\alpha)^2 \INT \bar{\psi}(\mathcal{D}x)\RD\g{\mu}A_{\mu}(x)\RD \psi(\mathcal{D}x)\\
		&=\I\INT\left[\bar{\psi}(x)\g{\mu}(\cos(\alpha)^2 A_{\mu}(x) +\sin(\alpha)^2 \AD )\psi(x)\right.\\
		&+\left.\I\sin(\alpha)\cos(\alpha)\,\bar{\psi}(x)\g{\mu}(A_{\mu}(x)
		-\AD)\RD\psi(\mathcal{D}x)\right],
	\end{split}
	\label{eq:intinv}
\end{equation}

\noindent
where in the 2nd equality we used the definition (\ref{eq:u1d_trans2}). 
Instead in the last equality we have done the same as for Eq. (\ref{eq:sfree}), changing the variable $x_{\mu}\TO x^{\mathcal{D}}_{\mu}=\mathcal{D}_{\mu\nu}x_{\nu}$, 
used Eq. (\ref{eq:rel1}) and the properties in Eq. (\ref{eq:rdprop}). 

As we observe, a sufficient condition for having the $U(1)_{\DD}$-invariance of $S_{\Int}$ is that 

\begin{equation}
	\AD = A_{\mu}(x), 
	\label{eq:acond}
\end{equation}

\noindent
since in that case, the 1st term of the last equality in (\ref{eq:intinv}) can be summed up to get the original $S_{\Int}$ and the 2nd term vanish. 
However, restricting ourself on only gauge configurations with $\AD = A_{\mu}(x)$, means to reject the configurations with non-zero topological charge, as we will see in a while. 
For proving so, we remember that the topological charge is defined as

\begin{equation}
	\nu(A) = \frac{1}{64\pi^2}\INT \epsilon_{\mu\nu\alpha\beta}F_{\mu\nu}(A;x)^a F_{\alpha\beta}(A;x)^a ,
	\label{eq:nua}
\end{equation}

\noindent
where $F_{\mu\nu}(A;x)^a = \partial^x_{\mu} A_{\nu}(x)^a -  \partial^x_{\nu} A_{\mu}(x)^a - gf_{abc}A_{\mu}(x)^b A_{\nu}(x)^c$ is the strength tensor
(keeping in mind that in non-abelian gauge theory, $A_{\mu}(x) = A_{\mu} (x)^a T^a$ and the $T^a$s are the generators of the gauge group, with $[ T^a ,T^b ] = \I f_{abc}T^c$). 
We observe that $F_{\mu\nu}(A^{\mathcal{D}};x)^a  = \mathcal{D}_{\mu\gamma}\mathcal{D}_{\nu\xi} F_{\gamma\xi}(A;\mathcal{D}x)^a $, where $F_{\gamma\xi}(A;\mathcal{D}x)^a  =  \partial^{\mathcal{D}x}_{\gamma} A_{\xi}(\mathcal{D}x)^a -  \partial^{^{\mathcal{D}x}}_{\xi} A_{\gamma}(\mathcal{D}x)^a -  gf_{abc}A_{\gamma}(\mathcal{D}x)^b A_{\xi}(\mathcal{D}x)^c$. 
Therefore the topological charge of $\AD$ is 

\begin{equation}
	\begin{split}
		&\nu(A^{\mathcal{D}})=
		\frac{1}{64\pi^2}\INT \epsilon_{\mu\nu\alpha\beta}F_{\mu\nu}(A^{\mathcal{D}};x)^a F_{\alpha\beta}(A^{\mathcal{D}};x)^a \\
		&= \frac{1}{64\pi^2}\INT \left[\epsilon_{\mu\nu\alpha\beta} 
		\mathcal{D}_{\mu\xi}
		\mathcal{D}_{\nu\omega}
		\mathcal{D}_{\alpha\gamma}
		\mathcal{D}_{\beta\lambda} F_{\xi\omega}(A;\mathcal{D}x)^a
		F_{\gamma\lambda}(A;\mathcal{D}x)^a\right]\\ 
		&= 
		\frac{1}{64\pi^2}\INTd \det(\mathcal{D}) \epsilon_{\mu\nu\alpha\beta}
		F_{\mu\nu}(A;x^{\mathcal{D}})^a
		F_{\alpha\beta}(A;x^{\mathcal{D}})^a \\ 
		&= \det(\mathcal{D})\nu(A) = -\nu(A),
	\end{split}
	\label{eq:nua2}
\end{equation}

\noindent
where we used that $\epsilon_{\mu\nu\alpha\beta} 
\mathcal{D}_{\mu\xi}
\mathcal{D}_{\nu\omega}
\mathcal{D}_{\alpha\gamma}
\mathcal{D}_{\beta\lambda}
=\det(\mathcal{D})\epsilon_{\xi\omega\gamma\lambda} 
$ which we can put outside the integral, and we have changed the variable $x_{\mu}\TO x^{\mathcal{D}}_{\mu}=\mathcal{D}_{\mu\nu}x_{\nu}$ exploiting that $\det(\mathcal{D}) = -1$, for $\mathcal{D} = \mathcal{P}$ or $\mathcal{T}$. 
Eq. (\ref{eq:nua2}) tells us that if $\AD = A_{\mu}(x)$, then $\nu(A) = \nu(A^{\mathcal{D}})= - \nu(A) = 0$. 
Hence gauge configurations with non-zero topological charge are not compatible with the condition (\ref{eq:acond}), that we have chosen for the invariance of $S_{\Int}$ in Eq. (\ref{eq:intinv}). 

For instance, let us look a special case of a gauge configuration for which $\nu(A)\neq 0 $ (and consequently the condition (\ref{eq:acond}) is not satisfied), and see how $U(1)_{\DD}$ invariance is broken in $S_{\Int}$. 
Considering the gauge group $SU(2)$, we can take the instanton configuration $A_{\mu}^{I}(x;\rho,\bar{x}) =  \eta^a_{\mu\nu}(x-\bar{x})_{\nu}\sigma^a/[(x-\bar{x})^2 + \rho^2]$, 
(for which, it is well-known that $\nu(A^{I})=1$, see Refs. \citen{Diakonov:2009jq,Belavin:1975fg}), placed in $\bar{x}$ and with size $\rho$, 
where $\sigma^a$ are the Pauli matrices in the color space and $\eta^a_{\mu\nu} = \epsilon_{a\mu\nu4} + \delta_{a\mu}\delta_{\nu4}-\delta_{a\nu}\delta_{\mu4}$ are the 't Hooft symbols. 
It is trivial to notice that $\eta^a_{\mu\nu}$ satisfies the property: 
$\mathcal{D}_{\mu\alpha}\eta^a_{\alpha\beta}\mathcal{D}_{\beta\nu} = \bar{\eta}^a_{\mu\nu}$ for $\mathcal{D} = \mathcal{P}$ or $\mathcal{T}$, 
where $\bar{\eta}^a_{\mu\nu} = \epsilon_{a\mu\nu4} - \delta_{a\mu}\delta_{\nu4}+\delta_{a\nu}\delta_{\mu4}$, which are the anti-self dual 't Hooft symbols. 
Now let us transform our instanton as $A_{\mu}^{I}(x;\rho,\bar{x})\TO A_{\mu}^{I}(x;\rho,\bar{x})^{\mathcal{D}}$, i.e. 

\begin{equation}
	\begin{split}
		&A_{\mu}^{I}(x;\rho,\bar{x})^{\mathcal{D}}= \mathcal{D}_{\mu\nu}A_{\nu}^{I}(\mathcal{D}x;\rho,\bar{x}) = \mathcal{D}_{\mu\alpha}
		\eta^a_{\alpha\nu}\frac{(\mathcal{D}x-\bar{x})_{\nu}\,\sigma^a}{[(\mathcal{D}x-\bar{x})^2 + \rho^2]}\\ 
		&= 
		\mathcal{D}_{\mu\alpha}
		\eta^a_{\alpha\nu}\mathcal{D}_{\nu\beta}\frac{(x-\mathcal{D}\bar{x})_{\beta}\,\sigma^a}{[(x-\mathcal{D}\bar{x})^2 + \rho^2]}= 
		\bar{\eta}^a_{\mu\nu}\frac{(x-\mathcal{D}\bar{x})_{\nu}\,\sigma^a}{[(x-\mathcal{D}\bar{x})^2 + \rho^2]}
		=A_{\mu}^{\bar{I}}(x;\rho,\mathcal{D}\bar{x}),
	\end{split}
	\label{eq:nua3}
\end{equation}

\noindent
where we used that $\mathcal{D}_{\mu\alpha}\mathcal{D}_{\alpha\nu}=\delta_{\mu\nu}$ 
from Eq. (\ref{eq:rdprop}) and therefore $(\mathcal{D}x-\bar{x})_{\alpha} = \mathcal{D}_{\alpha\beta}(x_{\beta} - \mathcal{D}_{\beta\gamma}\bar{x}_{\gamma}) = \mathcal{D}_{\alpha\beta}(x-\mathcal{D}\bar{x})_{\beta}$ and $(\mathcal{D}x-\bar{x})^2 = (\mathcal{D}x-\bar{x})_{\mu}(\mathcal{D}x-\bar{x})_{\mu} = 
\mathcal{D}_{\mu\alpha}\mathcal{D}_{\mu\beta}(x-\mathcal{D}\bar{x})_{\alpha}(x-\mathcal{D}\bar{x})_{\beta} = (x-\mathcal{D}\bar{x})^2$. 
Moreover, in the last equality of (\ref{eq:nua3}), we have changed to the anti self dual 't Hooft symbols. The expression after the 4th equality is referred to anti-instanton solution $A_{\mu}^{\bar{I}}(x;\rho,\mathcal{D}\bar{x})$, and it has topological charge $\nu(A^{\bar{I}})=-1$. This gauge field is placed at $\mathcal{D}\bar{x}$ and it has the same size of $A_{\mu}^{I}(x;\rho,\bar{x})$. 
As we expected the topological charge is therefore flipped from 1 to -1 and from Eq. (\ref{eq:intinv}) is evident that since $A_{\mu}^{I}(x;\rho,\bar{x}) \neq A_{\mu}^{\bar{I}}(x;\rho,\mathcal{D}\bar{x})$, then $S_{\Int}(\psi,\bar{\psi},A^{I})$ is not invariant under $U(1)_{\DD}$ transformations in such instanton gauge configuration. 

Finally, since we want to restrict to gauge configurations which satisfy (\ref{eq:acond}), and that consequently have zero topological charge,  we do not need to consider a possible 't Hooft term in the action (\ref{eq:sf2}), which arises from instanton configurations, since as we have seen, in such case, these possibilities already can break the fermionic action. 
\\

Therefore we conclude this section with Table \ref{tab:4}, in which the 2nd and 3rd column summarize the parts of action broken or invariant under $U(1)_{\DD}$ transformations.

		\begin{table}[htb]
			\tbl{Sectors of the QCD action which are invariant or broken by the groups $U(1)_{\DD}$, with $\Dlist$, and $SU(2)_{CS}^{\mathcal{D}}$ for $\mathcal{D} = \mathcal{P}$ or $\mathcal{T}$.}
				{\begin{tabular}{@{}cccc@{}}\toprule
					Action sector & $U(1)_{\PP}$, $U(1)_{\TT}$ & $U(1)_{\PA}$, $U(1)_{\TA}$ & $SU(2)_{CS}^{\mathcal{D}}$\\
					\colrule
					$S_F^0$ & invariant & invariant & invariant\\
					$S_m$ & invariant & broken & broken\\
					$S_{\Int}$ & invariant for $A_{\mu}= A_{\mu}^{\mathcal{D}}$ & invariant for $A_{\mu}= A_{\mu}^{\mathcal{D}}$ & invariant for $A_{\mu}= A_{\mu}^{\mathcal{D}}$\\
					\botrule
				\end{tabular}
				\label{tab:4}}
		\end{table}

\section{$SU(2)_{CS}$-like groups}\label{sec:su2cs0}

We are now at the stage of building two $SU(2)_{CS}$-like group transformations, 
which, as we will see, that they will look similar to the $SU(2)_{CS}$ group transformations of Refs. \citen{Denissenya:2014poa,Denissenya:2014ywa,Denissenya:2015mqa,Denissenya:2015woa}  and \citen{Glozman:2017dfd,Glozman:2016ayd} given in Eqs. (\ref{eq:su2cs_psi}) and (\ref{eq:su2cs_barpsi}). 
At first, we will examine their definition and the fermionic action invariance; then we will do some considerations regarding the change of reference frame; 
Finally we will see how in particular $SU(2)_{CS}^{\mathcal{P}}$ transformations (which we will define in the next section) are equivalent to the $SU(2)_{CS}$ ones in hadron temporal correlators.

\subsection{Definitions and invariance of the fermionic action}\label{sec:su2cs}

Such $SU(2)_{CS}$-like groups are obtained starting from the $U(1)_{\DD}$ transformations for $\Dlist$ and $U(1)_A$. 
In order to do this, we need to introduce a bit of notation. 
We start defining four fields for $\mathcal{D} = \mathcal{P}$ and $\mathcal{T}$, i.e. $\psi_{\pm}(x) = \frac{1}{2}(\psi(x)\pm\psi(\mathcal{D}x))$ and 
$\bar{\psi}_{\pm}(x) = \frac{1}{2}(\bar{\psi}(x)\pm\bar{\psi}(\mathcal{D}x))$, so that $\psi(x) = \psi_{+}(x) + \psi_{-}(x)$ and $\bar{\psi}(x) = \bar{\psi}_{+}(x) + \bar{\psi}_{-}(x)$, 
moreover $\psi_{\pm}(\mathcal{D}x) = \pm\psi_{\pm}(x)$ and $\bar{\psi}_{\pm}(\mathcal{D}x) = \pm\bar{\psi}_{\pm}(x)$. 
Now the $U(1)_A$ transformations for $\psi_{\pm}$ and $\bar{\psi}_{\pm}$ (given the ones for $\psi$ and $\bar{\psi}$) are easy to get and they are  
$\psi_{\pm}(x)\TO\psi_{\pm}(x)^{U_A^{\alpha}}=\exp(-\I\alpha\g{5})\psi_{\pm}(x)$ and
$\bar{\psi}_{\pm}(x)\TO\bar{\psi}_{\pm}(x)^{U_A^{\alpha}}=\bar{\psi}_{\pm}(x)\exp(-\I\alpha\g{5})$.
Regarding the $U(1)_{\DD}$ transformations of such new fields for $\Dlist$, we can just replace $\psi_{\pm}$ and $\bar{\psi}_{\pm}$ in the definition (\ref{eq:u1d_trans2}), 
instead of $\psi$ and $\bar{\psi}$ and use the previous considerations. 
In this case, we, therefore, obtain that $\psi_{\pm}(x)\TO\psi_{\pm}(x)^{U_{\DD}^{\alpha}} = \exp(\pm\I\alpha\RD) \psi_{\pm} (x)$ 
and $\bar{\psi}_{\pm}(x)\TO\bar{\psi}_{\pm}(x)^{U_{\DD}^{\alpha}} =  \bar{\psi}_{\pm} (x)\exp(\mp\I\eta\,\alpha\RD)$.

Furthermore we introduce the vectors  

\begin{equation}
	\PSI(x) = \left(\begin{matrix}
		\psi_{+} (x)\\
		\psi_{-} (x)
	\end{matrix}\right)\quad\mbox{and}\quad
	\PSIb(x) = \left(\begin{matrix}
		\bar{\psi}_{+} (x) &
		\bar{\psi}_{-} (x)
	\end{matrix}\right).
	\label{eq:spinor_new}
\end{equation}

\noindent
This permits us to rewrite the $U(1)_A$ transformations of $\psi_{\pm}$ and $\bar{\psi}_{\pm}$ as 

\begin{equation}
	\begin{split}
		U(1)_A:\quad&\PSI(x)\TO\Psi^{\mathcal{D}}(x)^{U_A^{\alpha}}=\exp(\I\alpha(-\mathds{1}_{\mathcal{D}}\otimes \g{5}))\,\PSI(x), \\
		&\PSIb(x)\TO\bar{\Psi}^{\mathcal{D}}(x)^{U_A^{\alpha}} = \PSIb(x)\,\exp(\I\alpha(-\mathds{1}_{\mathcal{D}}\otimes \g{5})) 
	\end{split}
	\label{eq:bbb}
\end{equation}

\noindent
and the $U(1)_{\DD}$ transformations for $\Dlist$ as 

\begin{equation}
	\begin{split}
		U(1)_{\DD}:\quad&\PSI(x)\TO\Psi^{\mathcal{D}}(x)^{U_{\DD}^{\alpha}}= \exp(\I\alpha(\sigma^3_{\mathcal{D}}\otimes\RD))\,\PSI(x),\\
		&\PSIb(x)\TO\bar{\Psi}^{\mathcal{D}}(x)^{U_{\DD}^{\alpha}}= \PSIb(x)\,\exp(-\I\eta\,\alpha(\sigma^3_{\mathcal{D}}\otimes\RD)),
	\end{split}
	\label{eq:bbb2}
\end{equation}

\noindent
where $\sigma^3_{\mathcal{D}}$ is the 3rd Pauli matrix and $\mathds{1}_{\mathcal{D}}$ is a $2\times 2$ identity and 
they act on the 2-dimensional space induced by Eq. (\ref{eq:spinor_new}) for a given $\mathcal{D} = \mathcal{P}$ or $\mathcal{T}$. 
At this point, we have found the right notation in order to define our $SU(2)_{CS}$-like groups, but at first we need to define their generators. 
For this purpose, we notice that the following set of matrices (for $\mathcal{D} = \mathcal{P}$ or $\mathcal{T}$) 

\begin{equation}
	\Sigma_i^{\mathcal{D}}=\{\sigma^3_{\mathcal{D}}\otimes\g{4},\sigma^3_{\mathcal{D}}\otimes\I\g{5}\g{4},-\mathds{1}_{\mathcal{D}}\otimes\g{5}\},
	\label{eq:su2cslike_gen}
\end{equation}

\noindent
forms an $su(2)$ Lie algebra, 
$\left[\Sigma_i^{\mathcal{D}},\Sigma_j^{\mathcal{D}}\right] = 2\I\epsilon_{ijk}\Sigma_k^{\mathcal{D}}$, they are hermitian, i.e. 
$\Sigma_i^{\mathcal{D}\,\dagger} = \Sigma_i^{\mathcal{D}}$ and traceless $\Tr(\Sigma_i^{\mathcal{D}})=0$, for $i=1,2,3$. 

We call the group generated by the matrices in (\ref{eq:su2cslike_gen}) as $SU(2)_{CS}^{\mathcal{D}}$ (for $\mathcal{D} = \mathcal{P}$ or $\mathcal{T}$) and we define the $SU(2)_{CS}^{\mathcal{D}}$ transformations of the fields $\Psi^{\mathcal{D}}$ and $\bar{\Psi}^{\mathcal{D}}$ as 

\begin{equation}
	\begin{split}
		SU(2)_{CS}^{\mathcal{D}}:\quad
		&\PSI(x)\TO\PSI(x)^{(\Sigma^{\mathcal{D}})}= 
		\exp(\I\alpha_n\Sigma_n^{\mathcal{D}})\PSI(x),\\
		&\PSIb(x)\TO\PSIb(x)^{(\Sigma^{\mathcal{D}})}= 
		\PSIb(x)(\mathds{1}_{\mathcal{D}}\otimes\g{4})\exp(-\I\alpha_n\Sigma_n^{\mathcal{D}})(\mathds{1}_{\mathcal{D}}\otimes\g{4}),
	\end{split}
	\label{eq:su2cslike_trans}
\end{equation}

\noindent
where $\bm{\alpha} = (\alpha_1,\alpha_2,\alpha_3)$ is a real global vector. 
Now, considering Eq. (\ref{eq:su2cslike_trans}), we observe that for different orientations of $\bm{\alpha}$, then $SU(2)_{CS}^{\mathcal{D}}$ has three different $U(1)$ subgroups. 
Namely from Eq. (\ref{eq:bbb}) and the expressions of $\RD$ and $\eta$ given in Table \ref{tab:2}, we have that $SU(2)^{\mathcal{P}}$ has the three subgroups: $U(1)_A$, $U(1)_{\PA}$ and $U(1)_{\PP}$; 
while $SU(2)^{\mathcal{T}}$ has the subgroups: $U(1)_A$, $U(1)_{\TT}$ and $U(1)_{\TA}$, which for both $SU(2)_{CS}^{\mathcal{D}}$ correspond to the three orthogonal orientations of $\bm{\alpha}$, i.e. $ (0,0,\alpha_3)$, $(0,\alpha_2,0)$ and $(\alpha_1,0,0)$ respectively. 
We summarize the information of such groups in Table \ref{tab:6}.

\begin{table}[htb]
\tbl{Main $U(1)$ subgroups of $SU(2)_{CS}^{\mathcal{D}}$ (for $\mathcal{D}=\mathcal{P}$ or $\mathcal{T}$) obtained from Eqs. (\ref{eq:bbb})-(\ref{eq:su2cslike_trans}), for different directions of $(\alpha_1,\alpha_2,\alpha_3)$.}
		{\begin{tabular}{@{}cccc@{}}\toprule
			$U(1)\subset SU(2)_{CS}^{\mathcal{D}}$ & Generator & Trans. for $\PSI$ & $\bm{\alpha}$\\
			\colrule
			$U(1)_{\PP}\subset SU(2)_{CS}^{\mathcal{P}}$ 	 & $\sigma^3_{\mathcal{P}}\otimes\g{4}$			& $\exp(\I\alpha(\sigma^3_{\mathcal{P}}\otimes\g{4}))$ & $(\alpha,0,0)$\\
			$U(1)_{\PA}\subset SU(2)_{CS}^{\mathcal{P}}$ & $\sigma^3_{\mathcal{P}}\otimes\I\g{5}\g{4}$	&  $\exp(\I\alpha(\sigma^3_{\mathcal{P}}\otimes\I\g{5}\g{4}))$ & $(0,\alpha,0)$\\
			$U(1)_{\TA}\subset SU(2)_{CS}^{\mathcal{T}}$ 	 & $\sigma^3_{\mathcal{T}}\otimes\g{4}$			& $\exp(\I\alpha(\sigma^3_{\mathcal{T}}\otimes\g{4}))$ & $(\alpha,0,0)$\\
			$U(1)_{\TT}\subset SU(2)_{CS}^{\mathcal{T}}$ & $\sigma^3_{\mathcal{T}}\otimes\I\g{5}\g{4}$	&  $\exp(\I\alpha(\sigma^3_{\mathcal{T}}\otimes\I\g{5}\g{4}))$ & $(0,\alpha,0)$\\
			$U(1)_A\subset SU(2)_{CS}^{\mathcal{D}}$ 	 & $-\mathds{1}_{\mathcal{D}}\otimes\g{5} $		&  $\exp(\I\alpha(-\mathds{1}_{\mathcal{D}}\otimes\g{5}))$ & $(0,0,\alpha)$\\
			\botrule
		\end{tabular}
		\label{tab:6}}
\end{table}

\noindent
As general fact, we point out that if we want to prove that a certain quantity (function of $\psi$ and $\bar{\psi}$) is invariant 
under $SU(2)_{CS}^{\mathcal{D}}$ transformations, we just need to prove that it is invariant under the three subgroups corresponding to the 
three generators of $SU(2)_{CS}^{\mathcal{D}}$. 
In our case, we have seen in Table \ref{tab:4}, that $S^0_F$ remains invariant under $U(1)_{\PP}$, $U(1)_{\TT}$, $U(1)_{\PA}$ and $U(1)_{\TA}$ transformations of $\psi$ and $\bar{\psi}$. 
It is actually also invariant under $U(1)_A$ transformations, since $S_F^0$ is the free massless fermionic action. 
Therefore we can conclude that $S_F^0$ is also invariant under both  transformations in Eq. (\ref{eq:su2cslike_trans}), for $\mathcal{D} = \mathcal{P}$ or $\mathcal{T}$ (however for the interested ones we report a more direct demonstration of the $SU(2)_{CS}^{\mathcal{D}}$ invariance of $S_F^0$ in Appendix \ref{app:E}). 
We have also seen in the previous section, that if we restrict to gauge configurations which satisfy the condition (\ref{eq:acond}), then $S_{\Int}$ is invariant under $U(1)_{\DD}$ group transformations as well. 
Using also that $S_{\Int}$ is $U(1)_A$-invariant, then $S_{\Int}$ is consequently invariant under the whole group  $SU(2)_{CS}^{\mathcal{D}}$ with our restriction of the gauge fields, i.e. $\AD = A_{\mu}(x)$. 
Everything is summarized in the last column of Table \ref{tab:4}.
Therefore we arrived to the conclusion that while $SU(2)_{CS}$ of section \ref{sec:old} does not leave the action of free massless fermions $S_F^0$ invariant, 
instead $SU(2)_{CS}^{\mathcal{D}}$ does it. 
This is the reason why a possible presence of the $SU(2)_{CS}^{\mathcal{D}}$ symmetry looks not in contrast with the deconfinement regime of QCD. 
However such possible presence of $SU(2)_{CS}^{\mathcal{D}}$ symmetry in bound states is something which need to be verified 
(for instance  experimentally or by lattice calculations) and at this stage the fact that $SU(2)_{CS}^{\mathcal{D}}$ is a symmetry of the free massless 
action does not imply that this symmetry should emerge in bound states at high temperature QCD. 
Nevertheless we will come back again in section \ref{sec:correlators} to this point.

\subsection{Translating the reference frame}\label{sec:inst}

In the above discussions we always assumed some fixed reference frame on which to apply our parity and time-reversal operation and therefore define $U(1)_{\DD}$ and consequently $SU(2)_{CS}^{\mathcal{D}}$ transformations. 
Now, we want to show what happens when we change reference frame. 
For instance, translating it. 

Suppose to translate our space-time points $x_{\mu}\TO x_{\mu}^a = x_{\mu} + a_{\mu}$ of a given vector $a_{\mu}$. 
In this case, defining the spinor fields $\psi^a(x^a)\equiv\psi(x+a)$ and 
$\bar{\psi}^a(x^a)\equiv\bar{\psi}(x+a)$, it is easy to show that $S_F^0 (\psi,\bar{\psi}) = S_F^0 (\psi^a,\bar{\psi}^a)$. 
Therefore, we can now apply the $U(1)_{\DD}$ and $SU(2)_{CS}^{\mathcal{D}}$ transformations on the new fields $\psi^a$ and $\bar{\psi}^a$, then $S_F^0$ still remains invariant for whatever $a_{\mu}$ we choose. 
We call the groups $U(1)_{\DD}$ or $SU(2)_{CS}^{\mathcal{D}}$ acting on
$\psi^a$ and $\bar{\psi}^a$ (instead of $\psi$ and $\bar{\psi}$) as 
$SU(2)_{CS}^{\mathcal{D},a}$ and $U(1)_{\DD,a}$. 
Therefore our original groups are just $U(1)_{\DD}\equiv U(1)_{\DD,a=0}$ and $SU(2)_{CS}^{\mathcal{D}} = SU(2)_{CS}^{\mathcal{D},a=0}$, 
but, as we said, the action of free massless fermions $S_F(\psi,\bar{\psi})$ is invariant for the groups $SU(2)_{CS}^{\mathcal{D},a}$ and $U(1)_{\DD,a}$, no matter what is the value of $a_{\mu}$.

However, concerning the gauge part of the action $S_{\Int}(\bar{\psi},\psi,A)$, which is of course invariant under space-time translations, this does not necessary mean that it is invariant under $SU(2)_{CS}^{\mathcal{D},a}$ and $U(1)_{\DD,a}$, for whatever $a_{\mu}$ we choose, even if $A_{\mu}$ satisfies the relation in (\ref{eq:acond}) in a given reference frame.
We clarify this in more details, giving an example.

Let us restrict to gauge fields $A_{\mu}$ which satisfies the condition in Eq. (\ref{eq:acond}) and therefore we know that $S_{\Int}(\psi,\bar{\psi},A)$ is invariant under $SU(2)_{CS}^{\mathcal{D},a=0}$ and $U(1)_{\DD,a=0}$ transformations.

For instance, a gauge field obeying the condition (\ref{eq:acond}) can be obtained just looking Eq. (\ref{eq:nua3}) of how on instanton field transforms under $\mathcal{D}$ operations. 
We can observe that the following combination 

\begin{equation}
	A_{\mu}^{I\bar{I}}(x;\rho) = A^{I}_{\mu}(x;\rho,\bar{x}) + A^{\bar{I}}_{\mu}(x;\rho,\mathcal{D}\bar{x}),
	\label{eq:mol1}
\end{equation}

\noindent
which is often regarded as instanton molecule (see Refs. \citen{Ilgenfritz:1988dh,Ilgenfritz:1994nt,Schafer:1994nv}), 
already satisfies the property in (\ref{eq:acond}), because 
$\mathcal{D}_{\mu\nu}A_{\nu}^{I\bar{I}}(\mathcal{D}x;\rho) = 
\mathcal{D}_{\mu\nu} A^{I}_{\nu}(\mathcal{D}x;\rho,\bar{x})  + \mathcal{D}_{\mu\nu}A^{\bar{I}}_{\nu}(\mathcal{D}x;\rho,\mathcal{D}\bar{x})= A^{\bar{I}}_{\mu}(x;\rho,\mathcal{D}\bar{x}) +  A^{I}_{\mu}(x;\rho,\bar{x}) = A_{\mu}^{I\bar{I}}(x;\rho) $ (from Eq. (\ref{eq:nua}), it implies that $\nu(A^{I\bar{I}})=0$). 
Consequently $S_{\Int}(\bar{\psi},\psi,A^{I\bar{I}})$ is invariant under $SU(2)_{CS}^{\mathcal{D},a=0}$ and $U(1)_{\DD,a=0}$ transformations.
However if we change reference frame (i.e. $x_{\mu} = x_{\mu}^a - a_{\mu}$) and we consider
the group transformations $SU(2)_{CS}^{\mathcal{D},a}$ and $U(1)_{\DD,a}$, the situation is pretty different. 
In that case the operator $\mathcal{D}$ acts on the new coordinates, i.e.  $\mathcal{D}_{\mu\nu}\,x^a_{\nu} = \mathcal{D}_{\mu\nu}\,x_{\nu} + \mathcal{D}_{\mu\nu}\,a_{\nu}$. 
The action on the gauge field in the new reference frame is also different. 
First of all, we rewrite Eq. (\ref{eq:mol1}) using the expression of the (anti-) instanton solution, that we saw in section \ref{subsec:int}, and we get that $A_{\mu}^{I\bar{I}}(x;\rho) =  A^{I}_{\mu}(x^a-a;\rho,\bar{x}) + A^{\bar{I}}_{\mu}(x^a-a;\rho,\mathcal{D}\bar{x}) =
A^{I}_{\mu}(x^a;\rho,\bar{x}+a) + A^{\bar{I}}_{\mu}(x^a;\rho,\mathcal{D}\bar{x}+a) $.
Therefore, it is better to define the new molecule in the new reference frame as 
\begin{equation}
	A_{\mu}^{I\bar{I}}(x^a;\rho)^a \equiv A^{I}_{\mu}(x^a;\rho,\bar{x}+a) + A^{\bar{I}}_{\mu}(x^a;\rho,\mathcal{D}\bar{x}+a),
	\label{eq:aref}
\end{equation}

\noindent
and as we see the positions are now translated of a vector $a_{\mu}$. 

At this point if we want to check the possible  $SU(2)_{CS}^{\mathcal{D},a}$ and $U(1)_{\DD,a}$ invariance of $S_{\Int}(\psi,\bar{\psi},A^{I\bar{I}})$, the fields has to satisfy the condition in Eq. (\ref{eq:acond}) in the new reference frame, which simply looks like $\mathcal{D}_{\mu\nu}A_{\nu}^{I\bar{I}}(\mathcal{D}x^a;\rho)^a =A_{\mu}^{I\bar{I}}(x^a;\rho)^a$. However is this really satisfied? Let us give a look at it: 

\begin{equation}
	\begin{split}
		\mathcal{D}_{\mu\nu}A_{\nu}^{I\bar{I}}(\mathcal{D}x^a;\rho)^a &= 
		\mathcal{D}_{\mu\nu}A^{I}_{\nu}(\mathcal{D}x^a;\rho,\bar{x}+a) + \mathcal{D}_{\mu\nu}A^{\bar{I}}_{\nu}(\mathcal{D}x^a;\rho,\mathcal{D}\bar{x}+a)\\
		&=A^{\bar{I}}_{\nu}(x^a;\rho,\mathcal{D}\bar{x}+\mathcal{D}a) 
		+A^{I}_{\nu}(x^a;\rho,\bar{x}+\mathcal{D}a),
	\end{split}
\end{equation}

\noindent
that is equal to $A_{\mu}^{I\bar{I}}(x^a;\rho)^a$ in Eq. (\ref{eq:aref}) if $\mathcal{D}a = a$. 
This means that in general the condition in Eq. (\ref{eq:acond}) is not satisfied, hence $S_{\Int}(\psi,\bar{\psi},A^{I\bar{I}})$ is not invariant under $SU(2)_{CS}^{\mathcal{D},a}$ 
and $U(1)_{\DD,a}$ transformations, except for proper choices of the vector $a_{\mu}$.

Now, If we take for instance $\mathcal{D} = \mathcal{T}$ in Eq. (\ref{eq:mol1}) and we restrict to translations only in the spatial part, which means having translations of a quantity $a_{\mu}^S \equiv (a_1,a_2,a_3,0)$, then the condition 
$\mathcal{T}a^S = a^S$ is always satisfied. This means that for $\mathcal{D} = \mathcal{T}$, $S_{\Int}(\psi,\bar{\psi},A^{I\bar{I}})$ is invariant under $SU(2)_{CS}^{\mathcal{T},a^S}$, $U(1)_{\TT,a^S}$ and $U(1)_{\TA,a^S}$ transformations. 
However $\mathcal{P}a^S = -a^S$, so we do not expect to have the same invariance for the groups $SU(2)_{CS}^{\mathcal{P},a^S}$, $U(1)_{\PP,a^S}$ and $U(1)_{\PA,a^S}$. 
Vice versa, if we take $\mathcal{D} = \mathcal{P}$, 
and we restrict on translations in the temporal direction, i.e. of a quantity $a_{\mu}^T \equiv (0,0,0,a_4)$, then the condition $\mathcal{P}a^T = a^T$ is satisfied and $S_{\Int}(\psi,\bar{\psi},A^{I\bar{I}})$ is invariant under $SU(2)_{CS}^{\mathcal{P},a^T}$, $U(1)_{\PP,a^T}$ and $U(1)_{\PA,a^T}$ transformations. However since $\mathcal{T}a^T = -a^T$, then 
$SU(2)_{CS}^{\mathcal{T},a^T}$, $U(1)_{\TT,a^T}$ and $U(1)_{\TA,a^T}$ are not necessary symmetries of $S_{\Int}(\psi,\bar{\psi},A^{I\bar{I}})$.\\

This situation can be of course applied also for generic systems of multiple instanton molecules, as for example

\begin{equation}
	A^{N}_{\mu}(x) = \sum_{i=0}^N \left(A^{I}_{\mu}(x;\rho^{(i)},\bar{x}^{(i)}) + A^{\bar{I}}_{\mu}(x;\rho^{(i)},\mathcal{D}\bar{x}^{(i)})\right),
	\label{eq:nmol}
\end{equation}

\noindent
where it is easy to see that the condition $\mathcal{D}a = a$ still can be applied for this \textit{ansatz} as well.

This brings us to the conclusion that the structure of the gauge configurations selects the possible values of $a_{\mu}$ for which $SU(2)_{CS}^{\mathcal{D},a}$ are symmetry groups of the action of  massless fermions. Therefore not all $SU(2)_{CS}^{\mathcal{D},a}$ groups leave invariant the action of massless fermions. 
Moreover some structures for which the action is invariant under $SU(2)_{CS}^{\mathcal{P},a}$ transformations can be not invariant 
for $SU(2)_{CS}^{\mathcal{T},a}$ transformations and vice versa. 
This means that the possibility of having $SU(2)_{CS}^{\mathcal{P},a}$ and $SU(2)_{CS}^{\mathcal{T},a}$ group symmetries really depends by the interaction of fermions with the gauge fields, even if these symmetries are both compatible with the deconfinement in QCD, 
in the sense that they leave the action of free fermions $S_F^0$ invariant.

We conclude this section also clarifying an important point.  All the argumentation on the gauge fields given in this section is purely for example purposes, 
because restricting on only given gauge configurations does not make so much sense since the gauge fields are not something fixed but they have fluctuations with probability proportional to $\exp(-S_F + ...)$. 
Therefore we can expect that even a small fluctuation could break our $SU(2)_{CS}^{\mathcal{D},a}$ symmetries. 
Hence more investigations of physical cases are needed in order to estimate the size of such breaking and evaluate when these symmetries could emerge.

\subsection{Temporal correlators and $SU(2)_{CS}^{\mathcal{P}}$}\label{sec:correlators}

We have still left as open question the problem of having $SU(2)_{CS}^{\mathcal{D}}$ symmetries in hadron bound states at high temperature QCD where it approaches to the deconfinement regime. 
In this section, we want to address this problem, specifically for the group $SU(2)_{CS}^{\mathcal{P}}$. 
For such purpose, we take the $SU(2)_{CS}^{\mathcal{P}}$ action on the fermion field, which is given in Eq. (\ref{eq:su2cslike_trans}), at the point $x^{(t)}\equiv (\bm{0},x_4)$, where the space coordinates are set to zero. 
In this point, $\psi(\mathcal{P}x^{(t)}) = \psi(x^{(t)})$ and $\bar{\psi}(\mathcal{P}x^{(t)}) = \bar{\psi}(x^{(t)})$, which implies that $\psi_{-} (x^{(t)}) = 0$ and $\bar{\psi}_{-} (x^{(t)}) = 0$, but $\psi_{+} (x^{(t)}) = \psi(x^{(t)})$ and $\bar{\psi}_{+} (x^{(t)}) = \bar{\psi}(x^{(t)})$. 
Therefore the fields $\Psi^{\mathcal{P}}$ and $\bar{\Psi}^{\mathcal{P}}$  have only the upper component non-zero, namely 

\begin{equation}
	\Psi^{\mathcal{P}}(x^{(t)}) = \left(\begin{matrix}
		\psi (x^{(t)})\\
		0
	\end{matrix}\right),\quad
	\bar{\Psi}^{\mathcal{P}}(x^{(t)}) = \left(\begin{matrix}
		\bar{\psi} (x^{(t)}) &
		0
	\end{matrix}\right),
	\label{eq:spinor_new2}
\end{equation}

\noindent
(look Eq. (\ref{eq:spinor_new})). 
Now, from Eq. (\ref{eq:su2cslike_trans}) we get the transformations of $\Psi^{\mathcal{P}}(x^{(t)})$ under $SU(2)_{CS}^{\mathcal{P}}$, which we can write explicitly as

\begin{equation}
	\begin{split}
		SU(2)_{CS}^{\mathcal{P}}:\,&\left(\begin{matrix}
			\psi(x^{(t)})\\
			0
		\end{matrix}\right)\TO
		\left(\begin{matrix}
			\psi_{+}(x^{(t)})^{(\Sigma^{\mathcal{P}})}\\
			\psi_{-}(x^{(t)})^{(\Sigma^{\mathcal{P}})}
		\end{matrix}\right)\\
		&=  (\cos(\alpha) + \I\sin(\alpha)\,e_n\Sigma_n^{\mathcal{P}})
		\left(\begin{matrix}
			\psi(x^{(t)})\\
			0
		\end{matrix}\right),
	\end{split}
	\label{eq:xttrans}
\end{equation}

\noindent
where we wrote $\bm{\alpha} = \alpha(e_1,e_2, e_3)$ with $\sum_i  e_i^2 = 1$.
Eq. (\ref{eq:xttrans}) can be in principle further simplified. 
From Eq. (\ref{eq:su2cslike_gen}), we see that the generators of $SU(2)_{CS}^{\mathcal{P}}$ are $\Sigma^{\mathcal{P}}_n = \{\sigma^3_{\mathcal{P}}\otimes\g{4},\sigma^3_{\mathcal{P}}\otimes\I\g{5}\g{4},-\mathds{1}_{\mathcal{P}}\otimes\g{5}\}$, 
but $(\mathds{1}_{\mathcal{P}}\otimes\g{5})\Psi^{\mathcal{P}}(x^{(t)}) = (\g{5}\psi(x^{(t)})\quad 0)^T$, moreover $(\sigma^3_{\mathcal{P}}\otimes\I\g{5}\g{4})\,\Psi^{\mathcal{P}}(x^{(t)}) = (\I\g{5}\g{4}\,\psi(x^{(t)})\quad 0)^T$ 
and $(\sigma^3_{\mathcal{P}}\otimes\g{4})\,\Psi^{\mathcal{P}}(x^{(t)}) = (\g{4}\,\psi(x^{(t)})\quad 0)^T$ . 
This because $\sigma^3_{\mathcal{P}} = \diag(\mathds{1},-\mathds{1})$ has effect only on the upper component of $\Psi^{\mathcal{P}}(x^{(t)})$, since the lowest is zero. 
This means that 

\begin{equation}
	\Sigma^{\mathcal{P}}_n \Psi^{\mathcal{P}}(x^{(t)}) = 
	\Sigma^{\mathcal{P}}_n
	\left(\begin{matrix}
		\psi(x^{(t)})\\
		0
	\end{matrix}\right)=
	\left(\begin{matrix}
		\Sigma_n \psi(x^{(t)})\\
		0
	\end{matrix}\right),
\end{equation}

\noindent
where $\Sigma_n$ are the generators of $SU(2)_{CS}$ given in Eq. (\ref{eq:su2cs_generator}). 
Hence, from Eq. (\ref{eq:xttrans}), we get  

\begin{equation}
	\begin{split}
		SU(2)_{CS}^{\mathcal{P}}:\,&\psi(x^{(t)})\TO\psi(x^{(t)})^{(\Sigma^{\mathcal{P}})}\\
		&= (\cos(\alpha) + \I\sin(\alpha)\,e_n\Sigma_n) \psi(x^{(t)})\\
		&= \exp(\I\alpha_n\Sigma_n)\psi(x^{(t)}),
	\end{split}
	\label{eq:xttrans2}
\end{equation}

\noindent
where we took only the upper components of Eq. (\ref{eq:xttrans}), because the lowest is zero (more explicitly  $\psi_{+}(x^{(t)})^{(\Sigma^{\mathcal{P}})}= \psi(x^{(t)})^{(\Sigma^{\mathcal{P}})}$ and $\psi_{-}(x^{(t)})^{(\Sigma^{\mathcal{P}})} = 0$). 
Eq. (\ref{eq:xttrans2}) is basically an $SU(2)_{CS}$ transformation as given in Eq. (\ref{eq:su2cs_psi}). 
We can repeat the same procedure for $\bar{\Psi}(x^{(t)})$ in Eq. (\ref{eq:su2cslike_trans}) and obtain that 

\begin{equation}
	\begin{split}
		SU(2)_{CS}^{\mathcal{P}}:\,&\bar{\psi}(x^{(t)})\TO\bar{\psi}(x^{(t)})^{(\Sigma^{\mathcal{P}})}\\
		&= \bar{\psi}(x^{(t)})\g{4}(\cos(\alpha) - \I\sin(\alpha)\,e_n\Sigma_n)\g{4}\\
		&  = 
		\bar{\psi}(x^{(t)})\g{4}\exp(-\I\alpha_n\Sigma_n)\g{4}
	\end{split}
	\label{eq:xttrans3}
\end{equation}

\noindent
which is again exactly the same transformation given in Eq. (\ref{eq:su2cs_barpsi}) for $SU(2)_{CS}$. 

Therefore we conclude that a $SU(2)_{CS}^{\mathcal{P}}$ transformation of $\psi(x^{(t)})$ and $\bar{\psi}(x^{(t)})$ is identical to a $SU(2)_{CS}$ transformation on the same fields, 
i.e. $\psi(x^{(t)})^{(\Sigma^{\mathcal{P}})} = \psi(x^{(t)})^{(\Sigma)}$ and 
$\bar{\psi}(x^{(t)})^{(\Sigma^{\mathcal{P}})} = \bar{\psi}(x^{(t)})^{(\Sigma)}$. 
Hence on the space-time point $x^{(t)}$ we can not distinguish which group transformation we are applying, whatever it is $SU(2)_{CS}^{\mathcal{P}}$ or the simple $SU(2)_{CS}$.

Let us now apply this point on hadron correlators. 
We start, for instance with baryons. 
Typically a baryon operator can be expressed as $O_{\mathcal{B}}(x) = \mathcal{B}_{ijk}\,\psi_i(x)\psi_j(x)\psi_k(x)$, where $i,j,k$ are generic indices which incorporate those of color and Dirac ones, and $\mathcal{B}_{ijk}$ is a tensor specifying the quantum numbers of the given baryon we are considering. 
Now a $SU(2)_{CS}$ transformation of the spinors $\psi_i (x)$ induces a transformation of $O_{\mathcal{B}}(x)$ as well, and consequently it becomes 

\begin{equation}
	\begin{split}
		SU(2)_{CS}:\,&O_{\mathcal{B}}(x)\TO O_{\mathcal{B}}(x)^{(\Sigma)}\\
		& = \mathcal{B}_{ijk}\,\psi_i(x)^{(\Sigma)}\psi_j(x)^{(\Sigma)}\psi_k(x)^{(\Sigma)}\\ 
		&= b_1\,O_{\mathcal{B}}(x) + b_2\,O_{\mathcal{B}\g{4}}(x)\\
		& + b_3\,O_{\mathcal{B}\g{5}}(x) + b_4\,O_{\mathcal{B}\g{4}\g{5}}(x),
	\end{split}
	\label{eq:11}
\end{equation}

\noindent
where $O_{\mathcal{B}\mathcal{X}}(x) = (\mathcal{B}\mathcal{X})_{ijk}\,\psi_i(x)\psi_j(x)\psi_k(x)$, with $\mathcal{X} = \g{4},\g{5},\g{4}\g{5}$, 
and $\{b_i\}$ are complex coefficients depending by the parameters $(\alpha_1,\alpha_2,\alpha_3)$ of the $SU(2)_{CS}$ transformations given in Eq. (\ref{eq:su2cs_psi}). 
The same can be obtained for the adjoint euclidean operators 
$\bar{O}_{\mathcal{B}}(x) = \bar{\mathcal{B}}_{ijk}\,\bar{\psi}_i(x)\bar{\psi}_j(x)\bar{\psi}_k(x)$, i.e. 

\begin{equation}
	\begin{split}
		SU(2)_{CS}:\,&\bar{O}_{\mathcal{B}}(x)\TO \bar{O}_{\mathcal{B}}(x)^{(\Sigma)}\\
		& = \bar{\mathcal{B}}_{ijk}\,\bar{\psi}_i(x)^{(\Sigma)}\bar{\psi}_j(x)^{(\Sigma)}\bar{\psi}_k(x)^{(\Sigma)}\\ 
		&= \bar{b}_1\,\bar{O}_{\mathcal{B}}(x) + \bar{b}_2\,\bar{O}_{\mathcal{B}\g{4}}(x) \\ &+\bar{b}_3\,\bar{O}_{\mathcal{B}\g{5}}(x) + \bar{b}_4\,\bar{O}_{B\g{4}\g{5}}(x),
	\end{split}
	\label{eq:22}
\end{equation}

\noindent
where we used the transformations in (\ref{eq:su2cs_barpsi}) and 
$\bar{O}_{\mathcal{B}\mathcal{X}}(x) = (\bar{\mathcal{B}}\mathcal{X})_{ijk}\,\bar{\psi}_i(x)\bar{\psi}_j(x)\bar{\psi}_k(x)$, with $\mathcal{X} = \g{4},\g{5},\g{4}\g{5}$. 
$\{\bar{b}_i\}$ are again complex coefficients depending by the parameters $(\alpha_1,\alpha_2,\alpha_3)$.

Now, if for a proper choice of $(\alpha_1,\alpha_2,\alpha_3)$, we have that $O_B(x)^{(\Sigma)}$ gives us a physical baryon bound state (some of them has been studied in Ref. \citen{Denissenya:2015woa}) and $SU(2)_{CS}$ is a symmetry of the theory (which seems to be on lattice simulation studies,\cite{Denissenya:2014poa,Denissenya:2014ywa,Denissenya:2015mqa,Denissenya:2015woa}) 
then we have to observe a degeneracy of the masses related to the baryons  $O_{\mathcal{B}}(x)$ and $ O_{\mathcal{B}}(x)^{(\Sigma)}$. 
Moreover the following temporal correlators has to be equal

\begin{equation}
	\begin{split}
		&C_{\mathcal{B}}(T) = \langle O_{\mathcal{B}}(x^{(t)} + T\hat{4})\,\bar{O}_{\mathcal{B}}(x^{(t)})\quad\mbox{and}\\
		&C_{\mathcal{B}}(T)^{(\Sigma)} = \langle O_{\mathcal{B}}(x^{(t)} + T\hat{4})^{(\Sigma)}\,\bar{O}_{\mathcal{B}}(x^{(t)})^{(\Sigma)}\rangle,
	\end{split}
	\label{eq:corr11}
\end{equation}

\noindent
from which we can get the ground state masses for large $T$, since  $C_{\mathcal{B}}(T)\underset{T\TO\infty}{\sim}\exp(-m_{\mathcal{B}} T)$ and $C_{\mathcal{B}}(T)^{(\Sigma)}\underset{T\TO\infty}{\sim}\exp(-m_{\mathcal{B}}^{(\Sigma)} T)$, with $m_{\mathcal{B}}$ the baryon mass described by $O_{\mathcal{B}}$ and $m_{\mathcal{B}}^{(\Sigma)}$ the one related to $O_{\mathcal{B}}^{(\Sigma)}$. 
Therefore $m_{\mathcal{B}}$ has to be equal to $m_{\mathcal{B}}^{(\Sigma)}$ if $SU(2)_{CS}$ is a symmetry of the theory. 
However we have seen that $\psi(x^{(t)})^{(\Sigma^{\mathcal{P}})} = \psi(x^{(t)})^{(\Sigma)}$ and 
$\bar{\psi}(x^{(t)})^{(\Sigma^{\mathcal{P}})} = \bar{\psi}(x^{(t)})^{(\Sigma)}$, 
therefore on such space-time point $x^{(t)}$, the transformations of the baryon operators given in Eqs. (\ref{eq:11}) and (\ref{eq:22}) are the same for $SU(2)_{CS}^{\mathcal{P}}$. This means that 
$O_{\mathcal{B}}(x^{(t)})^{(\Sigma)} = O_{\mathcal{B}}(x^{(t)})^{(\Sigma^{\mathcal{P}})}$ and 
$\bar{O}_{\mathcal{B}}(x^{(t)})^{(\Sigma)} = \bar{O}_{\mathcal{B}}(x^{(t)})^{(\Sigma^{\mathcal{P}})}$, where we denoted with the label $\Sigma^{\mathcal{P}}$ the operators obtained by  $SU(2)_{CS}^{\mathcal{P}}$ transformations of $\psi_i (x)$ and $\bar{\psi}_i (x)$. 
In conclusion $C_{\mathcal{B}}(T)$ and $C_{\mathcal{B}}(T)^{(\Sigma)}$ are also connected via $SU(2)_{CS}^{\mathcal{P}}$ group transformations as well, and not only $SU(2)_{CS}$, so the study of such correlators does not distinguish the two group transformations. 
Moreover, since they are the same, then the degeneracy of the masses 
$m_{\mathcal{B}}$ and to $m_{\mathcal{B}}^{(\Sigma)}$ can be either explained by $SU(2)_{CS}$ or $SU(2)_{CS}^{\mathcal{P}}$. 
However the first group is not compatible with the deconfinement at high temperature QCD, while the second one it is. 

Now, in the recent high temperature studies\cite{Rohrhofer:2017grg,Rohrhofer:2019qwq,Glozman:2017dfd} the possible degeneracy of the hadron masses has not be studied yet, but only the correlators. 
However not the temporal ones in Eqs. (\ref{eq:corr11}), but correlators which are includes non zero momentum states. 
For instance the following correlators

\begin{equation}
	\begin{split}
		&C'_{\mathcal{B}}(T) = \sum_{\bm{x}}\langle O_{\mathcal{B}}(x + T\hat{4})\,\bar{O}_{\mathcal{B}}(x)\rangle\quad\mbox{and}\\
		&C'_{\mathcal{B}}(T)^{(\Sigma)} = \sum_{\bm{x}}\langle O_{\mathcal{B}}(x + T\hat{4})^{(\Sigma)}\,\bar{O}_{\mathcal{B}}(x)^{(\Sigma)}\rangle,
	\end{split}
	\label{eq:corr22}
\end{equation}

\noindent
in the momentum space include also terms with $\bm{p}\neq 0$, and for this reason they are only sensitive to $SU(2)_{CS}$, but not $SU(2)_{CS}^{\mathcal{P}}$, because for generic $x$ these two group transformations are in general different.  
Unfortunately for $T> 3 T_c$, $SU(2)_{CS}$ seems to disappear in such correlators as it should be also for deconfinement reasons. 
However, the symmetry $SU(2)_{CS}^{\mathcal{P}}$ still could be present at temperatures $T> 3 T_c$, because it does not go in contrast with the presence of deconfinement in QCD in this regime, 
but of course this is just a necessary condition, hence experiments and lattice calculations should check this point.

The same argument of above can be applied in straightforward way to mesons as well. 
A generic meson operator is given by $O_{\mathcal{M}}(x) = \mathcal{M}_{ij}\psi_i(x)\bar{\psi}_{j}(x)$, where as before $\mathcal{M}_{ij}$ is a tensor which depends by the quantum numbers we choose for our meson. 
Now as in Eq. (\ref{eq:11}), we can write the $SU(2)_{CS}$ transformation of $O_M(x)$ as 

\begin{equation}
	\begin{split}
		SU(2)_{CS}:\,&O_{\mathcal{M}}(x)\TO O_{\mathcal{M}}(x)^{(\Sigma)} \\
		&= \mathcal{M}_{ij}\,\psi_i(x)^{(\Sigma)}\bar{\psi}_j(x)^{(\Sigma)}\\ 
		&= m_1\,O_{\mathcal{M}}(x) + m_2\,O_{\mathcal{M}\g{4}}(x)\\
		& + m_3\,O_{\mathcal{M}\g{5}}(x) + m_4\,O_{\mathcal{M}\g{4}\g{5}}(x)
	\end{split}
	\label{eq:33}
\end{equation}

\noindent
with $\{m_i\}$ complex coefficients depending by the values of $(\alpha_1,\alpha_2,\alpha_3)$ in the $SU(2)_{CS}$ transformations in Eqs. (\ref{eq:su2cs_psi}) and (\ref{eq:su2cs_barpsi}), and $O_{\mathcal{M}\mathcal{X}}(x) = (\mathcal{M}\mathcal{X})_{ij}\psi_i(x)\bar{\psi}_{j}(x)$. 
The same of expression of Eq. (\ref{eq:33}), can be obtained for the adjoint operator $\bar{O}_{\mathcal{M}}(x) = \bar{\mathcal{M}}_{ij}\bar{\psi}_i(x)\psi_{j}$.  
Therefore if we substitute the label $\mathcal{B}\TO \mathcal{M}$ in Eqs. (\ref{eq:corr11}) and (\ref{eq:corr22}), we can repeat the same considerations as we have done for baryons and find out that a possible degeneracy in meson masses connected via $SU(2)_{CS}$ symmetry can be also explained by $SU(2)_{CS}^{\mathcal{P}}$.

\section{Conclusions}\label{sec:conclusions}

Here we give a summary of the main results in this paper. 

\begin{romanlist}
	\item From parity and time-reversal symmetries, 
	we have defined four $U(1)$ group transformations, namely $U(1)_{\PP}$, $U(1)_{\TT}$, $U(1)_{\PA}$ and $U(1)_{\TA}$, summarized in Eq. (\ref{eq:u1d_trans2}) and Table \ref{tab:2}. 
	$U(1)_{\PP}$ and $U(1)_{\TT}$ are group symmetries of the free fermionic action, and $U(1)_{\PA}$ and $U(1)_{\TA}$ are group symmetries of the free massless fermionic action. 
	\item The generators of the groups $U(1)_{\PP}$, $U(1)_{\PA}$ and $U(1)_A$ permit to define the group $SU(2)_{CS}^{\mathcal{P}}$; 
	and from the generators of $U(1)_{\TT}$, $U(1)_{\TA}$ and $U(1)_A$ we can instead define the group $SU(2)_{CS}^{\mathcal{T}}$. 
	Both these two $SU(2)$ groups are symmetries of the free massless fermion action. 
	\item The introduction of a gauge field breaks in general $SU(2)_{CS}^{\mathcal{P}}$ and $SU(2)_{CS}^{\mathcal{T}}$ (as well as the groups $U(1)_{\PP}$, $U(1)_{\TT}$, $U(1)_{\PA}$ and $U(1)_{\TA}$), 
	unless particular restrictions on the gauge field are considered. For instance, the invariance is preserved for gauge fields satisfying the relation in Eq. (\ref{eq:acond}) and are therefore in the zero topological sector. 
	
	\item We can define the group transformations $SU(2)_{CS}^{\mathcal{P}}$ and $SU(2)_{CS}^{\mathcal{T}}$, in different reference frames which are space-time translations of a generic vector $a_{\mu}$, from a given one. We have called such group transformations as $SU(2)_{CS}^{\mathcal{P},a}$ and $SU(2)_{CS}^{\mathcal{T},a}$. 
	The action of free massless fermions is invariant under $SU(2)_{CS}^{\mathcal{P},a}$ and $SU(2)_{CS}^{\mathcal{T},a}$ for whatever value of $a_{\mu}$ that we choose. 
	However when we consider the gauge interactions, then the possible gauge structures, which perhaps preserve the $SU(2)_{CS}^{\mathcal{P},a}$ and $SU(2)_{CS}^{\mathcal{T},a}$ symmetries in one reference frame, could not do this in others. 
	Therefore not all $SU(2)_{CS}^{\mathcal{P},a}$ and $SU(2)_{CS}^{\mathcal{T},a}$ for any $a_{\mu}$ are symmetries of the full fermionic action, but only some of them depending by the gauge field structure and for particular choices of the vector $a_{\mu}$.
	For this purpose, we have seen the example of instanton molecules, 
	and the conditions which $a_{\mu}$ need to satisfy in this case in order to have the above symmetries. 
	
	\item We have seen that a possible degeneracy in the hadron masses which can be explained by the group $SU(2)_{CS}$ (as it seems to happens in truncated lattice QCD studies, see Refs. \citen{Glozman:2017dfd,Glozman:2016ayd}), it can be also explained by $SU(2)_{CS}^{\mathcal{P}}$ (introduced in this paper). 
	Concerning the possible degeneracy of hadron correlators the situation is quite different. The temporal correlators (given for example for baryons in Eq. (\ref{eq:corr11})) are sensitive to both $SU(2)_{CS}$ and $SU(2)_{CS}^{\mathcal{P}}$ symmetries and therefore we could not distinguish them. Nevertheless the space averaged correlators in Eq. (\ref{eq:corr22}) are only sensitive to $SU(2)_{CS}$, because, in there, we have the contribution of states with momentum $\bm{p} \neq 0$, which spoil $SU(2)_{CS}^{\mathcal{P}}$. 
	Therefore if we hypothesize (as it has been done in Refs. \citen{Glozman:2017dfd,Glozman:2016ayd}) that at high temperatures $T\gg T_c$ a mass degeneracy of hadrons connected by $SU(2)_{CS}$ exists, then we can detect a possible $SU(2)_{CS}^{\mathcal{P}}$ symmetry, just looking at the degeneracy of the temporal correlators.  Meanwhile the averaged correlators in Eq. (\ref{eq:corr22}) could be not degenerate in such case, if the states with momentum $\bm{p} \neq 0$ become too much relevant on such regime. 
	This point is crucial because, so far, the lattice studies at high temperature QCD,\cite{Rohrhofer:2017grg,Rohrhofer:2019qwq,Rohrhofer:2019qal} focused on correlators only sensitive to $SU(2)_{CS}$ (not so different from the ones in Eq. (\ref{eq:corr22})) and without studying the hadron masses, but in this paper we want to remark that it is also worth to check the possible presence of $SU(2)_{CS}^{\mathcal{P}}$ symmetry at high temperature QCD, which differently from $SU(2)_{CS}$ is not in contract with the presence of deconfinement and for this reason, it could perhaps still visible also at $T>3T_c$. 
	
\end{romanlist}

\section*{Acknowledgments}

I am grateful to L. Glozman who introduced me to this topic and I want to thank J. Verbaarschot for the helpful conversations when I was at SBU. Moreover, I am really thankful to M. Marinkovic for the support.

\appendix

\section{Gamma matrices}\label{app:A}

The representation used in this paper for the gamma matrices in euclidean space-time is given by

\begin{equation}
	\g{\mu} = \left(
	\begin{matrix}
		0 & \bar{\sigma}_{\mu}\\
		\sigma_{\mu} & 0\\
	\end{matrix}
	\right),\qquad
	\g{5} = \g{4}\g{1}\g{2}\g{3}=\left(\begin{matrix}
		-\mathds{1} & 0\\
		0 & \mathds{1}
	\end{matrix}\right),
	\label{eq:gamma}
\end{equation}

\noindent
which is the usual chiral representation.
In (\ref{eq:gamma}), we denoted $\sigma_{\mu} = (\mathds{1},\I\bm{\sigma})$, $\bar{\sigma}_{\mu} = (\mathds{1},-\I\bm{\sigma})$, for $\mu=1,2,3,4$; 
while $\bm{\sigma} = (\sigma_1,\sigma_2,\sigma_3)$ are the Pauli matrices. 
From Eq. (\ref{eq:gamma}), we get the main gamma matrices properties: $\{\g{\mu},\g{\nu}\} = 2\delta_{\mu\nu}\mathds{1}$, for all $\mu,\nu=1,2,3,4$; $\{\g{\mu},\g5\}=0$, $\Tr(\g{\mu}) = \Tr(\g{5}) = \Tr(\g{\mu}\g{5})=0$ and $\g{\mu}^{\dagger} = \g{\mu}$ for all $\mu=1,2,3,4$; $\g{5}^{\dagger} = \g{5}$ and $\g{5}^2 = \mathds{1}$.

\section{On $\mathcal{O}(4)$ Lorentz transformations}

In the following, we report some interesting properties coming from the combination of $\mathcal{O}(4)$ Lorentz transformations and the groups $SU(2)_{CS}$ and $U(1)_{\DD}$.

\subsection{Lorentz transformations and $SU(2)_{CS}$}\label{app:B}

Here, we want to show how from the $\mathcal{O}(4)$ symmetry of the fermionic action in euclidean space-time we can construct $SU(2)_{CS}$ transformations equivalent to the ones introduced in section \ref{sec:old} by the generators in Eq. (\ref{eq:su2cs_generator}). 

At first we recall some basics of $\mathcal{O}(4)$ Lorentz transformations. 
Given a generic transformation: $x_{\mu} \TO x_{\mu}^{\Lambda} = \Lambda_{\mu\nu}x_{\nu}$, with $\Lambda\in\mathcal{O}(4)$, it implies spinor transformation as well, namely 
$\psi(x)\TO\psi(x)^{\Lambda} = S(\Lambda)\psi(\Lambda x)$ and $\bar{\psi}(x)\TO \bar{\psi}(x)^{\Lambda} = \bar{\psi}(\Lambda x)S(\Lambda)^{-1}$. 
The matrix $S(\Lambda)$ satisfies a few properties which mainly come from the requirement of the invariance of the fermionic action 

\begin{equation}
	S_F(\psi,\bar{\psi},A) = \INT \bar{\psi}(x) \left(\gmu(\partial_{\mu}^x + \I A_{\mu}(x)) + m\right)\psi(x),
	\label{eq:sf}
\end{equation}

\noindent
i.e. $S_F(\psi^{\Lambda},\bar{\psi}^{\Lambda},A^{\Lambda}) = S_F(\psi,\bar{\psi},A)$, where $A_{\mu}(x)^{\Lambda} = \Lambda_{\mu\nu}A_{\nu}(\Lambda x)$.  
These properties are $S(\Lambda^{-1}) = S(\Lambda)^{-1}$ and 
$S(\Lambda)^{-1} \g{\mu} S(\Lambda) = \Lambda_{\mu\nu} \g{\nu}$. 
In which the last property is basically a rotation of the gamma matrices. 
In particular, if we take the following matrix 

\begin{equation}
	\bar{\Lambda}_{\mu\nu} = \left(\begin{matrix}
		0 & 1 & 0 & 0\\
		0 & 0 & 1 & 0\\
		0 & 0 & 0 & 1\\
		1 & 0 & 0 & 0\\
	\end{matrix}\right)\in\mathcal{O}(4),
	\label{eq:lambar}
\end{equation}

\noindent
we can see that $\bar{\Lambda}_{\mu\nu} \g{\nu} = \g{(\mu \mod 4)+1}$, hence 
$S(\bar{\Lambda})^{-1} \g{\mu} S(\bar{\Lambda}) =\g{(\mu \mod 4)+1}$, for all $\mu=1,2,3,4$. 
This is interesting if we start from $\mu=4$, because we can apply powers of $\bar{\Lambda}$, for obtaining the other gamma matrices, more specifically

\begin{equation}
	S(\bar{\Lambda}^k)^{-1} \g{4}  S(\bar{\Lambda}^k) = \bar{\Lambda}^k_{4\nu}\g{\nu} = \g{k}\qquad
	\mbox{for}\quad k=1,2,3.
	\label{eq:labbartrans}
\end{equation}

Another important feature always coming from the Lorentz invariance of the action in Eq. (\ref{eq:sf}) is that considering the set of variables $\{\psi,\bar{\psi},A\}$ or $\{\psi^{\Lambda},\bar{\psi}^{\Lambda},A^{\Lambda}\}$ does not make any physical difference. 
Let us see the implication of this for our $SU(2)_{CS}$ group transformations. 
We start inverting the $\mathcal{O}(4)$ transformations in the spinor fields as $\psi(x) = S(\Lambda)^{-1} \psi^{\Lambda}(\Lambda^{-1} x)^{\Lambda}$ and $\bar{\psi}(x) = \bar{\psi}(\Lambda^{-1} x)^{\Lambda} S(\Lambda)$, so we can rewrite Eqs. (\ref{eq:su2cs_psi}) and (\ref{eq:su2cs_barpsi}) as

	\begin{equation}
		\begin{split}
			\psi(x)^{(\Sigma)} &= 
			S(\Lambda)^{-1} S(\Lambda) \exp(\I \alpha_n\Sigma_n)S(\Lambda)^{-1} \psi(\Lambda^{-1} x)^{\Lambda}\\
			&= S(\Lambda)^{-1} \exp(\I \alpha_nS(\Lambda)\Sigma_n S(\Lambda)^{-1}) \psi(\Lambda^{-1} x)^{\Lambda},\\
			\bar{\psi}(x)^{(\Sigma)} &=\bar{\psi}(\Lambda^{-1} x)^{\Lambda} S(\Lambda)\g{4} S(\Lambda)^{-1}(S(\Lambda)\exp(-\I \alpha_n\Sigma_n)S(\Lambda)^{-1})S(\Lambda)\g{4} S(\Lambda)^{-1} S(\Lambda)\\
			&=\bar{\psi}(\Lambda^{-1} x)^{\Lambda} (\Lambda_{4\nu}\g{\nu})^{-1} \exp(-\I \alpha_nS(\Lambda)\Sigma_n S(\Lambda)^{-1}) (\Lambda_{4\nu}\g{\nu})^{-1} S(\Lambda),
		\end{split}
		\label{eq:a4}
	\end{equation}

\noindent
where we used that $S(\Lambda)S(\Lambda)^{-1} = \mathds{1}$ and $S(\Lambda)\g{4} S(\Lambda)^{-1} = (S(\Lambda)^{-1}\g{4} S(\Lambda))^{-1} = (\Lambda_{4\nu}\g{\nu})^{-1}$. 
Now we change the variable $\Lambda^{-1} x \TO x $ and Eq. (\ref{eq:a4}) becomes

\begin{equation}
	\begin{split}
		&S(\Lambda)\, \psi(\Lambda x)^{(\Sigma)} = \exp(\I \alpha_n S(\Lambda)\Sigma_n \St\,\psi(x)^{\Lambda},\\
		&\bar{\psi}(\Lambda x)^{(\Sigma)}S(\Lambda)^{-1}\\
		& = \bar{\psi}(x)^{\Lambda}\,
		(\Lambda_{4\nu}\g{\nu})^{-1}
		\exp(-\I \alpha_nS(\Lambda)\Sigma_n S(\Lambda)^{-1})\,
		(\Lambda_{4\nu}\g{\nu})^{-1}.
	\end{split}
	\label{eq:a5}
\end{equation}

\noindent
At this point, we notice that $S(\Lambda)\psi(\Lambda x)^{(\Sigma)}$ and $\bar{\psi}(\Lambda x)^{(\Sigma)}S(\Lambda)^{-1}$, 
in the previous equation, are the Lorentz transformations of the spinor fields $\psi(x)^{(\Sigma)}$ and $\bar{\psi}(x)^{(\Sigma)}$ respectively, 
which means

\begin{equation}
	\begin{split}
		& \left(\psi(x)^{(\Sigma)}\right)^{\Lambda}  = S(\Lambda) \psi(\Lambda x )^{(\Sigma)},\\
		&(\bar{\psi}(x)^{(\Sigma)})^{\Lambda} = \bar{\psi}(\Lambda x)^{(\Sigma)} S(\Lambda)^{-1}.
	\end{split}
	\label{eq:a6}
\end{equation}

\noindent
We can name now $\Sigma^{\Lambda}_n\equiv S(\Lambda)\Sigma_n S(\Lambda)^{-1} $ and $\gamma^{\Lambda}\equiv (\Lambda_{4\nu}\g{\nu})^{-1}$, 
and moreover using the fact that it does not matter if in the fermionic action $S_F$ we use the fields $\psi$ and $\bar{\psi}$ instead of $\psi^{\Lambda}$ and $\bar{\psi}^{\Lambda}$, 
then we can substitute in Eq. (\ref{eq:a5}) $\psi^{\Lambda}\TO \psi$ and $\bar{\psi}^{\Lambda}\TO \bar{\psi}$ and rewrite the Eq. (\ref{eq:a5}) as 

\begin{equation}
	\begin{split}
		&\psi(x)^{(\Sigma^{\Lambda})} \equiv\left(\psi(x)^{(\Sigma)}\right)^{\Lambda}\vert_{\psi^{\Lambda}\to\psi} = \exp(\I\alpha_n\Sigma_n^{\Lambda}) \psi(x),\\
		& \bar{\psi}(x)^{(\Sigma^{\Lambda})} \equiv (\bar{\psi}(x)^{(\Sigma)})^{\Lambda}\vert_{\bar{\psi}^{\Lambda}\to\bar{\psi}}  = \bar{\psi}(x)\,\gamma^{\Lambda} 
		\exp(-\I\alpha_n\Sigma_n^{\Lambda})
		\gamma^{\Lambda}.
	\end{split}
	\label{eq:a7}
\end{equation}

\noindent
where we introduced on the left sides the new notation.
The Eqs. in (\ref{eq:a7}) are the same of Eqs. (\ref{eq:su2cs_psi}) and (\ref{eq:su2cs_barpsi}), but here the generators are changed and they are 

\begin{equation}
	\begin{split}
		\Sigma^{\Lambda}_i = \{S(\Lambda)\g{4}S(\Lambda)^{-1}, S(\Lambda)\I\g{5}\g{4}\St, -\Sl \g{5}\St\}.
	\end{split}
	\label{eq:su2cs_gen_l}
\end{equation}

We have therefore constructed new transformations just using $SU(2)_{CS}$ and $\mathcal{O}(4)$ transformations.
We can now show that the matrices $\Sigma_i^{\Lambda}$s form an $su(2)$ algebra. 
Indeed, we observe that since $\Sl \St = \mathds{1}$, then we have $[\Sigma^{\Lambda}_i,\Sigma^{\Lambda}_j] = \Sl [\Sigma_i,\Sigma_j]\St = 2\I\epsilon_{ijk} \Sl \Sigma_k \St = 2\I\epsilon_{ijk}\Sigma^{\Lambda}_k$,  moreover $\Tr(\Sigma_n^{\Lambda}) = \Tr(S(\Lambda)\Sigma_n S(\Lambda)^{-1}) = \Tr(\Sigma_n)=0$ for all $n=1,2,3$. 
These are two basic properties of Lie generators, the closure and traceless one respectively. 
Furthermore we can show that the matrices $\Sigma_n^{\Lambda}$s are also hermitian, since also the $\Sigma_n$s are. 
Let us see in details this point.
At first we have that $
(S(\Lambda)\g{\mu}S(\Lambda)^{-1})^{\dagger} = 
(S(\Lambda^{-1})^{-1}\g{\mu}S(\Lambda^{-1}))^{\dagger} =
(\Lambda_{\mu\nu}^{-1} \g{\nu})^{\dagger} = \Lambda_{\mu\nu}^{-1\,*} \g{\nu}^{\dagger} = \Lambda_{\mu\nu}^{-1} \g{\nu}
$, since $\Lambda^{-1\,*} = \Lambda^{-1}$ because $\Lambda\in\mathcal{O}(4)$ 
and they are transformations of the space-time coordinates which can be regarded as real matrices.
Moreover, we can rewrite $\Lambda^{-1}_{\mu\nu}\g{\nu} = S(\Lambda)\g{\mu}S(\Lambda)^{-1}$, therefore we have 

\begin{equation}
	(S(\Lambda)\g{\mu}S(\Lambda)^{-1})^{\dagger} = (S(\Lambda)\g{\mu}S(\Lambda)^{-1}),
\end{equation}

\noindent
which for $\mu=4$, we get that the first element of (\ref{eq:su2cs_gen_l}), is hermitian, i.e. $\Sigma_1^{\Lambda\,\dagger} = \Sigma_1^{\Lambda}$.
Regarding the hermiticity of $\Sigma_3^{\Lambda}$, we have that from the definition of $\g{5}$ in (\ref{eq:gamma}) we can split it as product of the four gamma matrices, rotated by $S(\Lambda)$, in formulae

\begin{equation}
	\begin{split}
		&(S(\Lambda)\g{5}S(\Lambda)^{-1})^{\dagger}  = (S(\Lambda^{-1})^{-1}\g{4}S(\Lambda^{-1})
		S(\Lambda^{-1})^{-1}\\
		&\qquad\times\g{1}S(\Lambda^{-1})
		S(\Lambda^{-1})^{-1}\g{2}S(\Lambda^{-1})
		S(\Lambda^{-1})^{-1}\g{3}S(\Lambda^{-1}))^{\dagger}.
	\end{split}
\end{equation} 

\noindent
Therefore re-using Eq. (\ref{eq:gamma}), we get  $(S(\Lambda)\g{5}S(\Lambda)^{-1})^{\dagger} = 
(\Lambda^{-1}_{4\alpha}
\Lambda^{-1}_{1\beta}
\Lambda^{-1}_{2\gamma}
\Lambda^{-1}_{3\xi}\,
\g{\alpha}\g{\beta}\g{\gamma}\g{\xi})^{\dagger} 
= (\Lambda^{-1}_{4\alpha}
\Lambda^{-1}_{1\beta}
\Lambda^{-1}_{2\gamma}
\Lambda^{-1}_{3\xi}\,
\g{\xi}\g{\gamma}\g{\beta}\g{\alpha})$ 
and consequently the hermiticity of the gamma matrices brings us to 
$(S(\Lambda)\g{5}S(\Lambda)^{-1})^{\dagger} = S(\Lambda)\g{4}\g{1}\g{2}\g{3}S(\Lambda)^{-1}$, 
and finally we obtain

\begin{equation}
	(S(\Lambda)\g{5}S(\Lambda)^{-1})^{\dagger} = S(\Lambda)\g{5}S(\Lambda)^{-1}.
\end{equation}

\noindent
This means that $\Sigma_3^{\Lambda\dagger} = \Sigma_3^{\Lambda}$. 
Now noticing from (\ref{eq:su2cs_gen_l}) that $\Sigma_2^{\Lambda} = \I\Sigma_1^{\Lambda}\Sigma_3^{\Lambda}$, 
we can see that $\Sigma_2^{\Lambda\dagger} = -\I \Sigma_3^{\Lambda}\Sigma_1^{\Lambda} = \I \Sigma_1^{\Lambda}\Sigma_3^{\Lambda}=\Sigma_2^{\Lambda}$, 
hence also $\Sigma_2^{\Lambda}$ is hermitian. 
Therefore $\Sigma^{\Lambda\,\dagger}_n = \Sigma^{\Lambda}_n$ for all $n=1,2,3$. 
Consequently the generators in Eq. (\ref{eq:su2cs_gen_l}) generates an $SU(2)$ group, that we call as $SU(2)_{CS}^{\Lambda}$ ($\equiv SU(2)_{CS} \times \mathcal{O}(4)$). 
Furthermore, as we already said, for the interchangeability of $\psi^{\Lambda}\leftrightarrow\psi$ and $\bar{\psi}^{\Lambda}\leftrightarrow\bar{\psi}$ on the left sides of Eq. (\ref{eq:a7}) due to the $\mathcal{O}(4)$ Lorentz invariance of the fermionic action $S_F$, 
we conclude that $SU(2)_{CS}^{\Lambda}$ is equivalent to $SU(2)_{CS}$. 

A special case is $\Lambda = \bar{\Lambda}^k$ with $k=1,2,3$. 
In this case Eq. (\ref{eq:labbartrans}) tells us that $S(\bar{\Lambda}^k)\g{4}S(\bar{\Lambda}^k)^{-1} = \g{k}^{-1} =  \g{k}$, 
where the last equality can be proved observing the expression of the gamma matrices in Eq. (\ref{eq:gamma}). 

Furthermore from the definition (\ref{eq:gamma}) of $\g{5}$, we have

\begin{equation}
	\begin{split}
		&S(\bar{\Lambda})\g{5} = S(\bar{\Lambda})\g{4}S(\bar{\Lambda})^{-1} S(\bar{\Lambda})\g{1}S(\bar{\Lambda})^{-1} S(\bar{\Lambda})\g{2}S(\bar{\Lambda})^{-1}\\
		&\qquad\times S(\bar{\Lambda})\g{3}S(\bar{\Lambda})^{-1} 
		S(\bar{\Lambda})=
		\g{1}\g{2}\g{3}\g{4}S(\bar{\Lambda}) = -\g{5}S(\bar{\Lambda}),
	\end{split}
\end{equation}

\noindent 
where we used that $\g{\mu}^{-1}= \g{\mu}$ and that from (\ref{eq:lambar}) $S(\bar{\Lambda})^{-1}\g{\mu}S(\bar{\Lambda}) = \g{(\mu\mod 4) + 1}$. 
Furthermore, $S(\bar{\Lambda})^2\g{5} = -S(\bar{\Lambda})\g{5}S(\bar{\Lambda}) = \g{5}S(\bar{\Lambda})^2$ and $S(\bar{\Lambda})^3\g{5} = S(\bar{\Lambda})\g{5}S(\bar{\Lambda})^2 = -\g{5}S(\bar{\Lambda})^3$, 
using also the group property: $S(\bar{\Lambda})^k = S(\bar{\Lambda}^k)$, we get $S(\bar{\Lambda}^k)\g{5} = (-1)^k \g{5}S(\bar{\Lambda}^k)$ for $k=1,2,3$. 
Thus the last matrix in (\ref{eq:su2cs_gen_l}) in $\Lambda = \bar{\Lambda}^k$ becomes $-S(\bar{\Lambda}^k) \g{5} S(\bar{\Lambda}^k)^{-1} = (-1)^{k+1}\g{5}$ and the second is 
$S(\bar{\Lambda}^k)\I \g{5}\g{4} S(\bar{\Lambda}^k)^{-1} = \I S(\bar{\Lambda}^k)\g{5}S(\bar{\Lambda}^k)^{-1}S(\bar{\Lambda}^k)\g{4} S(\bar{\Lambda}^k)^{-1}  = (-1)^k \I\g{5}\g{k}$.

Therefore we get that the generators for $SU(2)_{CS}^{\bar{\Lambda}^k}$ are 

\begin{equation}
	\Sigma^{\bar{\Lambda}^k}_i = \{\g{k},(-1)^k\I\g{5}\g{k},(-1)^{k+1}\g{5}\},
	\label{eq:su2cs_gen_k}
\end{equation}

\noindent
Hence for $\Lambda = \bar{\Lambda}^k$, in the transformations (\ref{eq:a7}), we just need to substitute the generators with the ones in (\ref{eq:su2cs_gen_k}) and it is easy to check that  $\gamma^{\bar{\Lambda}^k} = \g{k}^{-1} =  \g{k}$.

\subsection{Lorentz transformations and $U(1)_D$}\label{app:B2}

We repeat now the same procedure of Appendix \ref{app:B}, but applied on the group transformations $U(1)_D$ given in Eq. (\ref{eq:u1d_trans2}). 
This means that we want to show how from $\mathcal{O}(4)$ Lorentz symmetry of the fermionic action in euclidean space-time, 
we can construct other $U(1)_{\DD}$ group transformations which are equivalent to the one of Eq. (\ref{eq:u1d_trans2}), looking how $\mathcal{D}$ and $\RD$ change upon $\mathcal{O}(4)$ Lorentz transformations. 

As given in Appendix \ref{app:B}, we use that we can write $\psi(x) = S(\Lambda)^{-1}\psi(\Lambda^{-1}x)^{\Lambda}$ and $\bar{\psi}(x) = \bar{\psi}(\Lambda^{-1}x)^{\Lambda}S(\Lambda)$, with $\Lambda\in\mathcal{O}(4)$. 
This implies, from Eq. (\ref{eq:u1d_trans2}), that

	\begin{equation}
		\begin{split}
			&\psi(x)^{U^{\alpha}_{\DD}}= \cos (\alpha)\,S(\Lambda)^{-1}\psi(\Lambda^{-1}x)^{\Lambda} +  \I\sin (\alpha)\, R_D (\mathcal{D}) S(\Lambda)^{-1}\psi(\Lambda^{-1}\mathcal{D}x)^{\Lambda},\\
			&\bar{\psi}(x)^{U^{\alpha}_{\DD}}=\cos(\alpha)\,\bar{\psi}(\Lambda^{-1}x)^{\Lambda}S(\Lambda) -\I\eta\: \sin (\alpha)\, \bar{\psi}(\Lambda^{-1}\mathcal{D}x)^{\Lambda}S(\Lambda)R_D (\mathcal{D}).
		\end{split}
		\label{eq:u1d_trans3}
	\end{equation}

Now we can insert inside Eq. (\ref{eq:u1d_trans3}) the identity $S(\Lambda)S(\Lambda)^{-1}=\mathds{1}$, and therefore rewrite (\ref{eq:u1d_trans3}) as

	\begin{equation}
		\begin{split}
			&S(\Lambda) \psi(\Lambda x)^{U_{\DD}^{\alpha}} = \cos(\alpha)\,\psi(x)^{\Lambda} + \I \sin(\alpha)\,S(\Lambda) \RD S(\Lambda)^{-1}\psi(\Lambda^{-1}\mathcal{D}\Lambda x)^{\Lambda},\\
			&\bar{\psi}(\Lambda x)^{U_{\DD}^{\alpha}}S(\Lambda)^{-1}=
			\cos(\alpha)\,\bar{\psi}(x)^{\Lambda} -\I\eta\sin(\alpha)\,\bar{\psi}(\Lambda^{-1}\mathcal{D}\Lambda x)^{\Lambda} S(\Lambda)\RD S(\Lambda)^{-1},
		\end{split}
		\label{eq:u1d_c}
	\end{equation}

\noindent
where we have changed the variable $\Lambda^{-1}x \TO x$. 
On the left side of Eq. (\ref{eq:u1d_c}), we recognize the $\mathcal{O}(4)$ Lorentz transformation of the fields of $\psi^{U_{\DD}^{\alpha}}$ and $\bar{\psi}^{U_{\DD}^{\alpha}}$, 
hence, as we have done in Eq. (\ref{eq:a6}), we can define the transformed fields as 

\begin{equation}
	\begin{split}
		&\psi(x)^{U_{\DD}^{\Lambda\,\alpha}}\equiv (\psi(x)^{U_{\DD}^{\alpha}})^{\Lambda} = S(\Lambda)\psi(\Lambda x)^{U_{\DD}^{\alpha}}, \\
		&\bar{\psi}(x)^{U_{\DD}^{\Lambda\,\alpha}}\equiv (\bar{\psi}(x)^{U_{\DD}^{\alpha}})^{\Lambda} =
		\bar{\psi}(\Lambda x)^{U_{\DD}^{\alpha}}S(\Lambda)^{-1}.
	\end{split}
	\label{eq:u1d_c1}
\end{equation}

Now using the fact that we can always substitute $\psi^{\Lambda}\TO\psi$ and $\bar{\psi}^{\Lambda}\TO\bar{\psi}$, since the fermionic action $S_F$ is invariant under the $\mathcal{O}(4)$ Lorentz transformations, then (\ref{eq:u1d_c}) can be rewritten as 

\begin{equation}
	\begin{split}
		&\psi(x)^{U_{\DD}^{\Lambda\,\alpha}} = \cos(\alpha)\,\psi(x) + \I\sin(\alpha)\,\RDl \psi(\mathcal{D}^{\Lambda} x),\\
		&\bar{\psi}(x)^{U_{\DD}^{\Lambda\,\alpha}} = \cos(\alpha)\,\bar{\psi}(x) 
		-\I\eta\sin(\alpha)\,\bar{\psi}(\mathcal{D}^{\Lambda}x)\RDl,
	\end{split}
	\label{eq:u1d_c2}
\end{equation}

\noindent
where we defined $\mathcal{D}^{\Lambda} = \Lambda^{-1}\mathcal{D}\Lambda $ and $\RDl = S(\Lambda)\RD S(\Lambda)^{-1} $. 
Now we call the group transformations in Eq. (\ref{eq:u1d_c2}) as $U(1)_{\DD}^{\Lambda}$ to differentiate from $U(1)_{\DD}$ given in Eq. (\ref{eq:u1d_trans2}). 
Such $U(1)_{\DD}^{\Lambda}$ group is still abelian and unitary,
since $\mathcal{D}^{\Lambda}$ and $\RDl$ satisfy the same properties in Eq. (\ref{eq:rdprop}) (you can just replace in there $\mathcal{D}\TO\mathcal{D}^{\Lambda}$ and $\RD\TO\RDl$ and use some properties of $S(\Lambda)$ described in Appendix \ref{app:B}). 
Hence from Appendix \ref{app:C} and \ref{app:D}, we can easily prove the closure property and unitarity (the argumentation in there does not change upon substitutions $\mathcal{D}\TO\mathcal{D}^{\Lambda}$ and $\RD\TO\RDl$).

Furthermore, as we have seen, the fact that we can interchange between $\psi\leftrightarrow\psi^{\Lambda}$ and  $\bar{\psi}\leftrightarrow\bar{\psi}^{\Lambda}$ in Eq. (\ref{eq:u1d_c2}), without  changing the action $S_F$, tells us that the  $U(1)_{\DD}^{\Lambda}$ group transformations are equivalent to the 
$U(1)_{\DD}$ ones.

A special case is for $\Lambda = \bar{\Lambda}^k$, with $k=1,2,3$ and $\bar{\Lambda}$ given in Eq. (\ref{eq:lambar}). 
In this situation, we have $\mathcal{D}^{\bar{\Lambda}^k} = (\bar{\Lambda}^{k})^{-1}\mathcal{D}\bar{\Lambda}^{k}$ and for $\mathcal{D} = \mathcal{P}$, we have $(\bar{\Lambda}^{k})^{-1}\mathcal{P}\bar{\Lambda}^{k} = \mathcal{P}^{(k)}$, which is the parity operator about the k-axis, namely 
$\mathcal{P}^{(k)}=2\delta_{\mu\nu}\delta_{k\nu} - \delta_{\mu\nu}$, for $\mathcal{D} = \mathcal{T}$, we have $(\bar{\Lambda}^{k})^{-1}\mathcal{T}\bar{\Lambda}^{k} =- (\bar{\Lambda}^{k})^{-1}\mathcal{P}\bar{\Lambda}^{k} = -\mathcal{P}^{(k)}=\mathcal{T}^{(k)}$, that is the time-reversal operator about the k-axis. 
The expression for $R_{\DD}(\mathcal{D}^{\bar{\Lambda}^k})$ is also quite straightforward to obtain. 
Basically we use that $R_{\DD}(\mathcal{D}^{\bar{\Lambda}^k}) = S(\bar{\Lambda}^k)\RD S(\bar{\Lambda}^k)^{-1}$ and Eq. (\ref{eq:labbartrans}), which tells us that $S(\bar{\Lambda}^k) \g{4}  S(\bar{\Lambda}^k)^{-1}=(S(\bar{\Lambda}^k)^{-1} \g{4}  S(\bar{\Lambda}^k))^{-1} =  \g{k}^{-1} = \g{k}$ and that 
$S(\bar{\Lambda}^k) \g{5}  S(\bar{\Lambda}^k)^{-1} = (-1)^k \g{5}$. 
We summarize the results in Table \ref{tab:3}.

\begin{table}[htb]
	\tbl{Ingredients for the $U(1)_{\DD}^{\bar{\Lambda}^k}$ group transformations in Eq. (\ref{eq:u1d_c2}).}
		{\begin{tabular}{@{}cccc@{}}\toprule
			$D$ & $\mathcal{D}^{\bar{\Lambda}^k}$ & $R_{\DD}(\mathcal{D}^{\bar{\Lambda}^k})$ & $\eta$ \\
			\colrule
			P & $\mathcal{P}^{(k)}$ & $\g{k}$ & $1$\\
			T & $\mathcal{T}^{(k)}$ & $(-1)^k\I\g{k}\g{5}$ & $1$ \\
			PA & $\mathcal{P}^{(k)}$ & $(-1)^k\I\g{5}\g{k}$ & $-1$ \\
			TA & $\mathcal{T}^{(k)}$ & $\g{k}$ & $-1$\\
			\botrule
		\end{tabular}
		\label{tab:3}}
\end{table}

\section{On $U(1)_{\DD}$ and $SU(2)_{CS}^{\mathcal{D}}$ transformations}

In this Appendix, we give some important features of the $U(1)_{\DD}$ and $SU(2)_{CS}^{\mathcal{D}}$ transformations.

\subsection{Closure property of the $U(1)_{\DD}$ transformations}\label{app:C}

Here, we prove that using the properties of $\RD$ and $\mathcal{D}$ in Eq. (\ref{eq:rdprop}), then the $U(1)_{\DD}$ transformations, defined in Eqs. (\ref{eq:u1d_trans}) and (\ref{eq:u1d_trans2}), satisfy the closure property given in Eq. (\ref{eq:group_prop}). 
In words, we show that two consecutive $U(1)_{\DD}$ transformations to a spinor field is still a $U(1)_{\DD}$ transformation. 

At first we define for convenience $f^{(n)}_{\PP,\TT} = 1$ and 
$f^{(n)}_{\PA,\TA} = (\I\g{5})^{k_n}$, with $k_n = 4+(n\mod 2)$.
Thus from Eq. (\ref{eq:u1d_trans}) we have for a generic $\Dlist$,  

\begin{equation}
	\begin{split}
		(\psi(x)^{U_{\DD}^{\alpha}})^{U_{\DD}^{\beta}}&= \sum_{n=0}^{\infty}\frac{(\I\beta)^n}{n!}f^{(n)}_{\DD}\left[\sum_{m=0}^{\infty}\frac{(\I \alpha)^m}{m!}f^{(m)}_{\DD}\psi(x)^{\mathcal{D}^m}\right]^{\mathcal{D}^n}\\ 
		&= 
		\cos(\beta)\,\left[\sum_{m=0}^{\infty}\frac{(\I \alpha)^m}{m!}f^{(m)}_{\DD}\psi(x)^{\mathcal{D}^m}\right]\\
		&+\I\sin(\beta)\,\RD\left[\sum_{m=0}^{\infty}\frac{(\I \alpha)^m}{m!}f^{(m)}_{\DD}\psi(\mathcal{D}x)^{\mathcal{D}^m}\right]\\
		&=\cos(\beta)\, \left[\cos(\alpha)\,\psi(x) + \I\sin(\alpha)\,\RD\psi(\mathcal{D}x)\right]\\
		&+\I\sin(\beta)\,\RD\left[\cos(\alpha)\,\psi(\mathcal{D}x) + \I\sin(\alpha)\,\RD\psi(x)\right]\\
		&=\cos(\alpha + \beta)\,\psi(x) + \I\sin(\alpha + \beta)\,\RD\psi(\mathcal{D}x) = \psi(x)^{U_{\DD}^{\alpha+\beta}}, 
	\end{split}
	\label{eq:d1}
\end{equation}

\noindent
where in the first three equalities we used Eqs. (\ref{eq:u1d_trans}) and (\ref{eq:u1d_trans2}) 
and the linearity of the parity and time-reversal operation for spinors. 
In fact, taking $\psi_1(x)$ and $\psi_2 (x)$, two generic and independent spinors, 
we always have $(\alpha_1 \psi_1(x) + \alpha_2 \psi_2 (x))^{\mathcal{D}} = 
\alpha_1 \psi_1(x)^{\mathcal{D}}+ \alpha_2 \psi_2 (x)^{\mathcal{D}}$, and $\psi_i(x)^{\mathcal{D}}$ are given in Eq. (\ref{eq:par_temp}) for $\mathcal{D}  =\mathcal{P},\mathcal{T}$.
In the last equality of (\ref{eq:d1}), we used Eq. (\ref{eq:rdprop}), in particular $R_{\DD}(\mathcal{D})^2 = \mathds{1}$ and some simple trigonometric relations. 
The same procedure of Eq. (\ref{eq:d1}) can be also applied to the field $\bar{\psi}$ and we can find that $(\bar{\psi}(x)^{U_{\DD}^{\alpha}})^{U_{\DD}^{\beta}} = \bar{\psi}(x)^{U_{\DD}^{\alpha+\beta}}$.

\subsection{Unitarity of $U(1)_{\DD}$}\label{app:D}

In this subsection, we show
the unitarity of the $U(1)_{\DD}$ transformations.  
For doing so, we only need to prove that 
exists a scalar product of the spinor fields which is left invariant under $U(1)_{\DD}$ transformations.
For this end, we take the scalar product $(\psi_1,\psi_2)=\INT \psi_1(x)^{\dagger}\psi_2(x)$, 
between two generic spinor fields and see that it is invariant under $U(1)_{\DD}$ transformations of $\psi_1$ and $\psi_2$. 
We will just exploit the properties of $\RD$ and $\mathcal{D}$ in Eq. (\ref{eq:rdprop}). 
So let us take such scalar product calculated in $\psi_1(x)^{U_{\DD}^{\alpha}}$ and $\psi_2(x)^{U_{\DD}^{\alpha}}$, namely

\begin{equation}
	\begin{split}
		(\psi_1^{U_{\DD}^{\alpha}},\psi_2^{U_{\DD}^{\alpha}})&= 
		\INT \left[(\cos(\alpha)\,\psi_1(x) + \I\sin(\alpha)\,R_{\DD}(\mathcal{D})\psi_1(\mathcal{D}x))^{\dagger}\right.\\
		&\qquad\qquad\quad\left.(\cos(\alpha)\,\psi_2(x) + \I\sin(\alpha)\,R_{\DD}(\mathcal{D})\psi_2(\mathcal{D}x))\right]\\
		&=\cos(\alpha)^2\left[\INT \psi_1(x)^{\dagger}\psi_2(x)\right]\\  
		&+ \I\sin(\alpha)\cos(\alpha)\left[\INT \psi_1(x)^{\dagger}\RD\psi_2(\mathcal{D}x)\right] \\
		&-\I \sin(\alpha)\cos(\alpha)\left[\INT \psi_1(\mathcal{D}x)^{\dagger}\RD^{\dagger}\psi_2(x)\right]\\
		&+\sin(\alpha)^2\left[\INT \psi_1(\mathcal{D}x)^{\dagger}\RD^{\dagger}\RD\psi_2(\mathcal{D}x)\right]\\
		&=\cos(\alpha)^2\left[\INT \psi_1(x)^{\dagger}\psi_2(x)\right]\\ 
		&+\sin(\alpha)^2 \left[\INT \psi_1(x)^{\dagger}\RD^{\dagger}\RD\psi_2(x)\right] = (\psi_1,\psi_2),
	\end{split}
	\label{eq:d2}
\end{equation}

\noindent
where in the 1st and 2nd equalities, we used the definition (\ref{eq:u1d_trans2}); in the third one we performed the transformation $x_{\mu}\TO x_{\mu}^{\mathcal{D}}=\mathcal{D}_{\mu\nu}x_{\nu}$ and used that $d^4 x = \vert\det(\mathcal{D})\vert d^4 x^{\mathcal{D}} =  d^4 x^{\mathcal{D}}$, because $\det(\mathcal{D})=-1$, hence the terms proportional to $\sin(\alpha)\cos(\alpha)$ simplify to zero. 
In the last equalities we used that $\RD = \RD^{\dagger}$ and $\RD^2 = \mathds{1}$ for all $\Dlist$. 
The same argument used for Eq. (\ref{eq:d2}), can be trivially used to prove that $(\bar{\psi}_1^{\dagger},\bar{\psi}_2^{\dagger}) = ((\bar{\psi}_1^{U_{\DD}^{\alpha}})^{\dagger},(\bar{\psi}_2^{U_{\DD}^{\alpha}})^{\dagger})$, for generic and independent $\bar{\psi}_1$ and $\bar{\psi}_2$.

\subsection{$SU(2)_{CS}^{\mathcal{D}}$ and invariance of the free fermion action}\label{app:E} 

Here we prove that the action of free massless fermions $S_F^0 (\psi,\bar{\psi})$ is invariant under $SU(2)_{CS}^{\mathcal{D}}$ transformations given in Eq. (\ref{eq:su2cslike_trans}). 
For doing so, we rewrite $S_F^0 (\psi,\bar{\psi})$, given in (\ref{eq:sf0}), 
in terms of $\PSI(x)$ and $\PSIb(x)$ defined in Eq. (\ref{eq:spinor_new}), and for brevity but also more clarity we will omit the label $\mathcal{D}$, (but the reader has to keep in mind it). 
Hence we get

\begin{equation}
	\begin{split}
		&S_F^0 (\psi,\bar{\psi}) = \INT\bar{\psi}(x)\g{\mu}\partial_{\mu}^x \psi(x)\\
		& = \INT \left[\bar{\psi}_{+}(x)+\bar{\psi}_{-}(x)\right]\g{\mu}\partial_{\mu}^x \left[\psi_{+}(x)+\psi_{-}(x)\right]\\
		&=\frac{1}{2}\INT \bar{\Psi}(x)\left[\left(\mathds{1}+\sigma^1\right)\otimes\mathds{1}\right]
		(\mathds{1}\otimes\g{\mu})\left[\left(\mathds{1}+\sigma^1\right)\otimes\mathds{1}\right]\partial_{\mu}^x\Psi(x)\\
		&=\INT \bar{\Psi}(x)\left[\left(\mathds{1}+\sigma^1\right)\otimes\g{\mu}\right]\partial_{\mu}^x \Psi(x).
	\end{split}
	\label{eq:e1}
\end{equation}

\noindent
Therefore the $SU(2)_{CS}^{\mathcal{D}}$ transformation of $S_F^0 (\psi,\bar{\psi})$ looks like 

\begin{equation}
	\begin{split}
		&S_F^0 (\psi^{(\Sigma^{\mathcal{D}})},\bar{\psi}^{(\Sigma^{\mathcal{D}})})\\
		&= \INT	\bar{\Psi}(x)^{(\Sigma^{\mathcal{D}})}\left[\left(\mathds{1}+\sigma^1\right)\otimes\g{\mu}\right]\partial_{\mu}^x \Psi(x)^{(\Sigma^{\mathcal{D}})}\\
		&=\INT \bar{\Psi}(x)(\mathds{1}\otimes\g{4})e^{-\I\alpha_n\Sigma_n^{\mathcal{D}}}(\mathds{1}\otimes\g{4})\left[\left(\mathds{1}+\sigma^1\right)\otimes\g{\mu}\right]e^{\I\alpha_m\Sigma_m^{\mathcal{D}}}\partial_{\mu}^x \Psi(x),
	\end{split}
	\label{eq:e2}
\end{equation}

\noindent
where we have substituted the $SU(2)_{CS}^{\mathcal{D}}$ group transformations of $\Psi(x)$ and $\bar{\Psi}(x)$, given in Eq. (\ref{eq:su2cslike_trans}), 
and used that, since the parameters $\alpha_i$s do not depend by $x$, we can exchange with $\partial_{\mu}^x$. Thus the structure of (\ref{eq:e2}) is given by 

\begin{equation}
	S_F^0 (\psi^{(\Sigma^{\mathcal{D}})},\bar{\psi}^{(\Sigma^{\mathcal{D}})}) = \INT \bar{\Psi}(x) \Gamma_{\mu}\partial_{\mu}^x \Psi(x),
	\label{eq:e3}
\end{equation}

\noindent
where, using that 	$\Sigma_i^{\mathcal{D}}=\{\sigma^3\otimes\g{4},\sigma^3\otimes\I\g{5}\g{4},-\mathds{1}\otimes\g{5}\}$ (see Eq. (\ref{eq:su2cslike_gen})), we have denoted

\begin{equation}
	\begin{split}
		\Gamma_{\mu} &= 
		(\mathds{1}\otimes\g{4})\left\lbrace\cos(\alpha) - \I\sin(\alpha)\left[e_1 (\sigma^3\otimes\g{4})\right.\right.\\ 
		&\left.\left.+ e_2(\sigma^3\otimes\I\g{5}\g{4}) + e_3(-\mathds{1}\otimes\g{5})\right] \right\rbrace(\mathds{1}\otimes\g{4})\\
		&\times\left[\left(\mathds{1}+\sigma^1\right)\otimes\g{\mu}\right]
		\left\lbrace\cos(\alpha) + \I\sin(\alpha)\right.\\
		&\times\left.\left[e_1 (\sigma^3\otimes\g{4}) + e_2(\sigma^3\otimes\I\g{5}\g{4}) + e_3(-\mathds{1}\otimes\g{5})\right] \right\rbrace,
	\end{split}
	\label{eq:e4}
\end{equation}

\noindent
in which we expanded $\exp(\I\alpha_n\Sigma_n^{\mathcal{D}}) = \cos(\alpha) + \I\sin(\alpha)\, e_n \Sigma_n^{\mathcal{D}}$ (the sum over $n$ is understood), 
with $(\alpha_1,\alpha_2,\alpha_3) = \alpha(e_1,e_2,e_3)$ and $\sum_{i=1}^3 e_i^2 = 1$. 
Now we use the property of $\Psi(x)$ and $\bar{\Psi}(x)$, which comes from the definition in (\ref{eq:spinor_new}), i.e. $\Psi(x)= (\sigma^3\otimes\mathds{1})\Psi(\mathcal{D}x)$ and $\bar{\Psi}(x) = \bar{\Psi}(\mathcal{D}x)(\sigma^3\otimes\mathds{1})$ and expanding $\Gamma_{\mu}$ in all of its terms, i.e.  $\Gamma_{\mu} = \sum_i g_{\mu}^{(i)}$ and rewrite (\ref{eq:e3}) as 

\begin{equation}
	\begin{split}
		&S_F^0 (\psi^{(\Sigma^{\mathcal{D}})},\bar{\psi}^{(\Sigma^{\mathcal{D}})}) = \sum_i\INT \bar{\Psi}(x) g_{\mu}^{(i)}\partial_{\mu}^x \Psi(x)\\
		&= \sum_i\INT \bar{\Psi}(\mathcal{D}x)(\sigma^3\otimes\mathds{1}) g_{\mu}^{(i)}(\sigma^3\otimes\mathds{1})\partial_{\mu}^x \Psi(\mathcal{D}x)\\
		&= \sum_i\INT \bar{\Psi}(x)\left[(\sigma^3\otimes\mathds{1}) g_{\nu}^{(i)}(\sigma^3\otimes\mathds{1})\mathcal{D}_{\nu\mu}\partial_{\mu}^x\right] \Psi(x),\\
		\label{eq:e5}
	\end{split}
\end{equation}

\noindent
where in the 3rd equality we have changed the variable $x\to\mathcal{D}x$ and used that the Jacobian $\vert\det(\mathcal{D})\vert=1$. 
From (\ref{eq:e5}) we notice that under the ``sandwich'' with $\bar{\Psi}(x)$ and $\Psi(x)$ and then integration $\INT$, we have that $g_{\mu}^{(i)}$ and $[(\sigma^3 \otimes\mathds{1})g_{\nu}^{(i)}(\sigma^3\otimes\mathds{1})\mathcal{D}_{\nu\mu}]$ can be interchanged without any effect on $S_F^0 (\psi^{(\Sigma^{\mathcal{D}})},\bar{\psi}^{(\Sigma^{\mathcal{D}})})$. 
In this sense, we consider them equivalent, and we write symbolically:

\begin{equation}
	g_{\mu}^{(i)} \stackrel{\int}{=} 
	(\sigma^3 \otimes\mathds{1})g_{\nu}^{(i)}(\sigma^3\otimes\mathds{1})\mathcal{D}_{\nu\mu},
	\label{eq:e6}
\end{equation}

\noindent
where with the symbol $``\stackrel{\int}{=} " $ we remark that they are equivalent and it can be substituted with $``="$ if on both sides of (\ref{eq:e6}) we multiply by $\bar{\Psi}(x)$ and $\Psi(x)$ on the right and left respectively and then we integrate with $\INT$, as for example is given  in Eq. (\ref{eq:e5}) (look first and last line). 

Now let us expand $\Gamma_{\mu}$ in (\ref{eq:e4}) and use (\ref{eq:e6}), 

	\begin{equation}
		\begin{split}
			\Gamma_{\mu} &= \cos(\alpha)^2 [(\mathds{1}+\sigma^1)\otimes\g{\mu}] + \I\sin(\alpha)\cos(\alpha)\left[ e_1[(\mathds{1}+\sigma^1)\otimes\g{\mu}(\sigma^3\otimes\g{4})]\right.\\ &\left.+e_2[(\mathds{1}+\sigma^1)\otimes\g{\mu}(\sigma^3\otimes\I\g{5}\g{4})] + e_3[(\mathds{1}+\sigma^1)\otimes\g{\mu}(-\mathds{1}\otimes\g{5})]\right]\\
			&-\I\sin(\alpha)\cos(\alpha)\left[e_1 [(\sigma^3\otimes\g{4})(\mathds{1}+\sigma^1)\otimes\g{\mu}] - e_2[(\sigma^3 \otimes\I\g{5}\g{4})(\mathds{1}+\sigma^1)\otimes\g{\mu}]\right.\\
			&\left. -e_3 [(-\mathds{1}\otimes\g{5})(\mathds{1}+\sigma^1)\otimes\g{\mu}]\right]+\sin(\alpha)^2\left[e_1^2[(\sigma^3\otimes\g{4})(\mathds{1}+\sigma^1)\otimes\g{\mu}(\sigma^3\otimes\g{4})]\right.\\
			&\left. +e_1 e_2 [(\sigma^3\otimes\g{4})(\mathds{1}+\sigma^1)\otimes\g{\mu}(\sigma^3\otimes\I\g{5}\g{4})] + e_1 e_3 [(\sigma^3\otimes\g{4})(\mathds{1}+\sigma^1)\otimes\g{\mu}(-\mathds{1}\otimes\g{5})]\right.\\
			&\left. -e_2 e_1 [(\sigma^3 \otimes\I\g{5}\g{4})(\mathds{1}+\sigma^1)\otimes\g{\mu}(\sigma^3\otimes\g{4})]-e_2^2[(\sigma^3\otimes\I\g{5}\g{4})(\mathds{1}+\sigma^1)\otimes\g{\mu}(\sigma^3\otimes\I\g{5}\g{4})]\right.\\
			&\left. -e_2 e_3 [(\sigma^3\otimes\I\g{5}\g{4})(\mathds{1}+\sigma^1)\otimes\g{\mu}(-\mathds{1}\otimes\g{5})] -e_3 e_1 [(-\mathds{1}\otimes\g{5})(\mathds{1}+\sigma^1)\otimes\g{\mu}(\sigma^3\otimes\g{4})]\right.\\
			&\left.-e_3 e_2 [(-\mathds{1}\otimes\g{5})(\mathds{1}+\sigma^1)\otimes\g{\mu}(\sigma^3\otimes\I\g{5}\g{4})] - e_3^2 (-\mathds{1}\otimes\g{5})(\mathds{1}+\sigma^1)\otimes\g{\mu}(-\mathds{1}\otimes\g{5})\right]\\
			&\stackrel{\int}{=} 
			\cos(\alpha)^2 [(\mathds{1}+\sigma^1)\otimes\g{\mu}] + \sin(\alpha)^2 \left(e_1^2
			+e_2^2 
			+e_3^2  \right)[(\mathds{1}+\sigma^1)\otimes\g{\mu}]=[(\mathds{1}+\sigma^1)\otimes\g{\mu}],
		\end{split}
		\label{eq:e7}
	\end{equation}

\noindent
where after $\stackrel{\int}{=}$ we used Eq. (\ref{eq:e6}) only for some terms $g_{\mu}^{(i)}$ in the expansion in order to simplify $\Gamma_{\mu}$. 
Using the linearity of the integral we plug (\ref{eq:e7}) in (\ref{eq:e3}) and we obtain

\begin{equation}
	S_F^0 (\psi^{(\Sigma^{\mathcal{D}})},\bar{\psi}^{(\Sigma^{\mathcal{D}})}) = \INT \bar{\Psi}(x) \left[(\mathds{1}+\sigma^1)\otimes\g{\mu}\right]\partial_{\mu}^x \Psi(x),
	\label{eq:e8}
\end{equation}

\noindent
that coincides with the right part of $(\ref{eq:e1})$, therefore we proved that $S_F^0 (\psi^{(\Sigma^{\mathcal{D}})},\bar{\psi}^{(\Sigma^{\mathcal{D}})})=
S_F^0 (\psi,\bar{\psi})$, which is the invariance of the action of free massless fermion under $SU(2)_{CS}^{\mathcal{D}}$ group transformations.

\bibliographystyle{ws-ijmpa}
\bibliography{chiralspin_catillo_2021_v2}

\end{document}